\documentclass[lettersize,journal]{IEEEtran}
\usepackage[T1]{fontenc}
\usepackage[utf8]{inputenc}
\usepackage{verbatim}
\usepackage{algorithm2e}
\usepackage{amsmath}
\usepackage{amsthm}
\usepackage{amssymb}
\usepackage{enumerate}
\usepackage{balance}

\theoremstyle{plain}
\newtheorem{thm}{\protect\theoremname}
\theoremstyle{plain}
\newtheorem{prop}[thm]{\protect\propositionname}

\usepackage{array}
\usepackage{multirow}
\SetAlFnt{\small}
\SetAlCapFnt{\small}
\SetAlCapNameFnt{\small}
\usepackage{bm}
\usepackage{steinmetz}
\usepackage[acronym,shortcuts]{glossaries}
\usepackage{todonotes}
\setuptodonotes{inline}
\usepackage{subfigure}
\usepackage{siunitx}

\theoremstyle{plain}
\providecommand{\factname}{Fact}
\theoremstyle{remark}
\theoremstyle{plain}

\usepackage{amsfonts}
\usepackage{cite}
\usepackage{array}

\providecommand{\lemmaname}{Lemma}
\providecommand{\theoremname}{Theorem}
\providecommand{\propositionname}{Proposition}
\usepackage{tikz}
\usepackage{pgfplots}

\usepackage{tabularx}
\usetikzlibrary{spy,arrows}
\usepackage{caption}

\usetikzlibrary{plotmarks,positioning}
\usetikzlibrary{spy,shapes,patterns,trees}
\usepackage{array}
\usetikzlibrary{backgrounds}
\tikzset{mark options={line width=1pt,solid}, font=\footnotesize}%

\usepackage{tikz}
\usepackage{pgf}\usepackage{pgfplots}
\usetikzlibrary{automata,positioning}
\usetikzlibrary{arrows,backgrounds,plotmarks}
\pgfplotsset{plot coordinates/math parser=false}
\usepackage{multicol}
\usepackage{lipsum}\usepackage{stfloats}
\tikzset{every picture/.style={line width=0.75pt}} 

\newacronym{$k$-NN}{$k$-NN}{k-Nearest Neighbors}
\newacronym{LSTM}{LSTM}{long short-term memory}
\newacronym[plural=RNNs,firstplural= recurrent neural networks (RNNs)]{RNN}{RNN}{recurrent neural network}
\newacronym[plural=CFRs,firstplural= channel frequency responses (CFRs)]{CFR}{CFR}{channel frequency response}
\newacronym{PLA}{PLA}{physical layer authentication}
\newacronym[plural=AoAs,firstplural= angle of arrivals (AoAs)]{AoA}{AoA}{angle of arrival}
\newacronym{MIMO}{MIMO}{multiple-input multiple-output}
\newacronym{SNR}{SNR}{signal-to-noise ratio}
\newacronym{ULA}{ULA}{uniform linear array}

\makeatother

\providecommand{\propositionname}{Proposition}
\providecommand{\theoremname}{Theorem}

\begin{document}

\title{Leveraging Angle of Arrival Estimation against Impersonation Attacks in Physical Layer Authentication} 

\author{Thuy M. Pham,~\IEEEmembership{Member,~IEEE}, 
    Linda Senigagliesi,~\IEEEmembership{Member,~IEEE}, 
    Marco Baldi,~\IEEEmembership{Senior Member,~IEEE},
    Rafael F. Schaefer,~\IEEEmembership{Senior Member,~IEEE},
    Gerhard P. Fettweis,~\IEEEmembership{Fellow,~IEEE},
    and Arsenia Chorti,~\IEEEmembership{Senior Member,~IEEE},

\thanks{This work is financed by the Saxon State government out of the State budget approved by the Saxon State Parliament. M. Baldi is supported by project SERICS (PE00000014) under the MUR National Recovery and Resilience Plan, funded by the European Union - Next Generation EU. L. Senigagliesi and A. Chorti have been partially supported by the EC through the Horizon Europe/JU SNS project ROBUST-6G (Grant Agreement no. 101139068) and the EU HORIZON MSCA-SE TRACE-V2X project (Grant No. 101131204). A. Chorti has been partially supported by the ANR-PEPR 5G Future Networks project, the CYU INEX-PHEBE and NESLI project. R. F. Schaefer and G. P. Fettweis have further been supported in part by the German Federal Ministry of Education and Research (BMBF) through the research hub 6G-life under Grant 16KISK001K and in part by the German Research Foundation (DFG) as part of Germany’s Excellence Strategy – EXC 2050/1 - Project ID 390696704 -- Cluster of Excellence \emph{``Centre for Tactile Internet with Human-in-the-Loop'' (CeTI)}. This publication is also based upon work from COST Action 6G-PHYSEC (CA22168), supported by COST (European Cooperation in Science and Technology).}
\thanks{An initial study on this approach was presented at the 2023 IEEE Global Communications Conference (GLOBECOM 2023), Kuala Lumpur, Malaysia, 4-8 December 2023  \cite{Pham2023}.}
\thanks{
Thuy M. Pham is with the Wireless Connectivity
and Sensing Group, Barkhausen Institut, 01067 Dresden, Germany (e-mail: minhthuy.pham@barkhauseninstitut.org).}
\thanks{L. Senigagliesi and A. Chorti are with the ETIS, UMR 8051 CY Cergy Paris University, ENSEA, CNRS, 95000 Cergy, France (e-mail: \{linda.senigagliesi, arsenia.chorti\}@ensea.fr).}
\thanks{Marco Baldi is with the Department of Information Engineering, Università Politecnica delle Marche, 60131 Ancona, Italy, and also with CNIT, 43124 Parma, Italy (e-mail: m.baldi@staff.univpm.it).}
\thanks{
R. F. Schaefer and G. P. Fettweis are with the Technische Universit\"at Dresden, the BMBF Research Hub 6G-life, the Cluster of Excellence \emph{``Centre for Tactile Internet with Human-in-the-Loop (CeTI),''} 01062 Dresden, Germany, and the Barkhausen Institut, 01067 Dresden, Germany (e-mail: \{rafael.schaefer, gerhard.fettweis\}@tu-dresden.de).}
}

\maketitle
 
\begin{abstract}
In this paper, we investigate the utilization of the angle of arrival (AoA) as a feature for robust physical layer authentication (PLA). While most of the existing approaches to PLA focus on common features of the physical layer of communication channels, such as channel frequency response, channel impulse response or received signal strength, the use of AoA in this domain has not yet been studied in depth, particularly regarding the ability to thwart impersonation attacks. In this work, we demonstrate that an impersonation attack targeting AoA-based PLA is only feasible under strict conditions on the attacker’s location and hardware capabilities, which highlights the AoA's potential as a strong feature for PLA.
We extend previous works considering a single-antenna attacker to the case of a multiple-antenna attacker, and we develop a theoretical characterization of the conditions in which a successful impersonation attack can be mounted.
Furthermore, we leverage extensive simulations in support of theoretical analyses, to validate the robustness of AoA-based PLA.   
\end{abstract}

\begin{IEEEkeywords}
Physical layer authentication, angle of arrival, antenna array, channel frequency response, channel impulse response, impersonation attack.
\end{IEEEkeywords}

\section{Introduction}
\label{sec:intro}
The extensive use of resource-constrained Internet of Things (IoT) devices in 5G and beyond presents notable security challenges. Traditional upper-layer authentication methods that rely on cryptography incur substantial overhead and latency, making them less suitable for this kind of applications. As a consequence, a lightweight authentication approach such as physical layer authentication (PLA), which leverages the unique and random characteristics of the physical properties for authentication, is gaining attention for sixth generation (6G) systems and networks \cite{Chorti:Contextaware:PLS:2022}.
In addition, by not requiring any assumption about the computational capacity of opponents, the PLA security guarantees are not affected by possible breakthroughs in the computational capacity of attackers, such as the availability of a sufficiently large quantum computer.

In general, PLA can be categorized into main two categories: device-based authentication and channel-based authentication. The former relies on hardware fingerprints, such as physically unclonable functions (PUFs) and impairments like I-Q imbalances, to be exploited as unique identifiers for devices \cite{Chorti23, Mitev22}. For instance, in a PUF-based authentication scheme, the operation relies on a challenge-response mechanism, in which any challenge generates a unique response, forming a challenge-response pair \cite{Pappu:PUF:2022,Gassend:SilionPUF:2002}. This scheme's security thus depends on the manufacturing process's complexity and controllability. An I-Q imbalance-based authentication exploits the mismatch between in-phase (I) and quadrature (Q) components which is common in wireless transceivers \cite{Hao:IQauthentication:2014,Hao:RelayIQ:2014}. These device-specific imperfections, though unique and uncontrollable, are also difficult to measure in practice. 

In contrast, channel-based PLA utilizes various properties of the communication channel, such as channel state information (CSI) and received signal strength (RSS), for authentication \cite{Chorti22}. RSS-based schemes commonly rely on statistical channel information and statistical tests, reconciliation \cite{Passah24}, or correlation computation to authenticate users \cite{Faria:Signalprints:2006,Xiao:EnhanceAuthentication:2008}.
Unlike RSS, which is a rather coarse feature, CSI - including channel frequency response (CFR) and channel impulse response (CIR) - provides instantaneous and richer channel information, and thus achieves better authentication performance \cite{Liu:CIR:PLA:2011,Liu:2Dquantization:PLA:2013,Xiao:CFR:PLA:2008, Senigagliesi21}. 
More recently, some studies in channel-based PLA have explored the angle of arrival (AoA) as a potential feature for authentication, besides CFR and CIR. For instance, the authors in \cite{Xiong:Securearray:2013} exploited AoA data to create a unique signature for each user in the system. In \cite{Abdelaziz2019}, an authentication scheme for vehicular communications was developed that calculates the expected AoA of the received signal based on reported GPS coordinates, which is then cross-verified with the estimated AoA. Furthermore, in \cite{Xu2022} hypothesis testing is employed to differentiate a legitimate base station from a rogue one based on the AoA of the received signal. Other works have applied similar authentication methods to low earth orbit (LEO) satellite constellations \cite{Topal2022} and underwater communications \cite{Casari2022}.

The use of AoA spectrum as a signature for authentication
was experimentally validated  in \cite{Xiong:Securearray:2013}. {\color{black}However, the authors did not consider any possible attacks to demonstrate the effectiveness of the method}. In \cite{Li:Proofing:Channel:AoA:2021}, a physical layer spoofing attack detection method is proposed, in which the AoA is included in the virtual channel representation. The authors in \cite{Abdelaziz:Sec:AoA:2016} focus on the robustness of the AoA against jamming attacks. However, the study of spoofing attacks on AoA-based authentication remains an open problem.  
For impersonation attacks, it has been shown in the literature that the AoA, when analog \ac{MIMO} array is used, is not a robust enough feature \cite{Murali24}. In this paper, we focus on digital array systems and demonstrate that they offer robustness against such attacks when the AoA is used as an authentication feature.

\subsection{Contribution}

In this paper, we investigate angle of arrival (AoA)-based PLA under spoofing attacks, in which an attacker seeks to impersonate a legitimate user. Specifically, we focus on examining potential impersonation attacks and thus establishing a condition under which such an impersonation attack can be executed. Our results show that such an attack is only feasible when the AoA of a single-antenna adversary is identical to that of the legitimate user.
Our analysis also shows that providing the attacker with more than one antenna does not actually produce any significant advantage in the execution of the attack.

In our initial investigation \cite{Pham2023}, we considered the simple case in which the adversary has only one antenna and only tries to manipulate the phase shift. In this  paper, we first generalize the single-antenna adversary case by taking into account a generic precoding factor that can change both phase and amplitude. Then, we extend the study to the two-antenna case and to the general case in which the attacker is provided with an arbitrary number of antennas. 
Our main contributions are thus summarized as follows:
\begin{itemize}
    \item We generalize the attacker model by considering a complex precoding matrix. We first consider a single-antenna attacker and then extend it to a multiple-antenna attacker.
    \item Conditions for a successful impersonation attack are established, according to which an attacker needs to be located at specific locations to forge an AoA identical to that of the legitimate user. 
    \item We numerically evaluate the performance of the system under different scenarios and show that numerical results are consistent with the theoretical analysis.
 \end{itemize}
 
\subsection{Notation}

Throughout the paper, bold lower- and upper-case letters represent vectors and matrices, respectively. $Re\{\cdot\}$ stands for the real part of a signal, $\mathbb{E}(\cdot)$ denotes the expectation of a random variable; $(\cdot)^{H}$ and $(\cdot)^{*}$ denote the Hermitian and conjugate operations, respectively. 

\subsection{Paper organization}

The rest of the paper is organized as follows. In Section \ref{sec:model}, we describe the system model under consideration, while in Section \ref{sec:AoA} we prove the condition for impersonation attacks utilizing AoA information. We then present numerical results in Section \ref{sec:num}. Finally, Section \ref{sec:con} concludes the paper.

\section{System model}

\label{sec:model}

In this section, we begin by recalling some 
fundamentals concerning AoA estimation and explain how it can be used
for PLA. 
We examine a standard authentication protocol where a receiver (Bob) must identify a legitimate transmitter (Alice) using the estimated \ac{AoA}, in presence of one (or more) opponent(s), whom we will refer to as Eve. The authentication process consists of two stages: an enrollment phase and a verification phase, structured as described below.
\begin{itemize}
    \item \textit{Enrollment phase:} Bob collects features, in our specific case estimated AoAs, belonging to different legitimate users. Channel estimation is usually made by assuming the transmission of pilot signals or messages with known content. In other words, during this phase, the authenticator is supposed to be able to map the different features to the identities of all users or nodes that may later request authentication. Note that, during this phase, Alice's transmissions to Bob are guaranteed to be authentic by upper-layer protocols.
    \item \textit{Verification phase:} Bob receives new messages from unknown transmitters and, based on the data collected in the previous stage, decides whether or not to authenticate them.
\end{itemize}

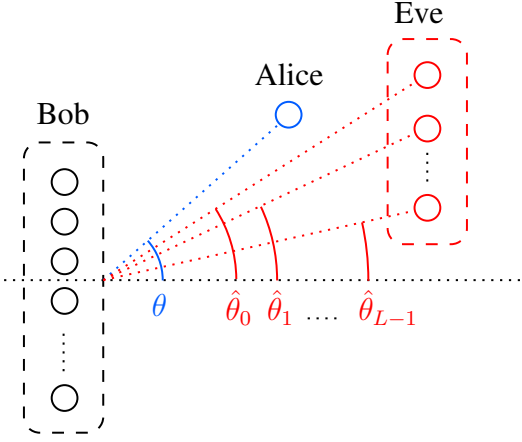
\begin{figure}
\begin{center}
\begin{tikzpicture}[x=0.75pt,y=0.75pt,yscale=-1,xscale=1]

\draw   (95.33,120.5) .. controls (95.33,116.91) and (98.24,114) .. (101.83,114) .. controls (105.42,114) and (108.33,116.91) .. (108.33,120.5) .. controls (108.33,124.09) and (105.42,127) .. (101.83,127) .. controls (98.24,127) and (95.33,124.09) .. (95.33,120.5) -- cycle ;
\draw   (95.33,140.5) .. controls (95.33,136.91) and (98.24,134) .. (101.83,134) .. controls (105.42,134) and (108.33,136.91) .. (108.33,140.5) .. controls (108.33,144.09) and (105.42,147) .. (101.83,147) .. controls (98.24,147) and (95.33,144.09) .. (95.33,140.5) -- cycle ;
\draw   (95.33,160.5) .. controls (95.33,156.91) and (98.24,154) .. (101.83,154) .. controls (105.42,154) and (108.33,156.91) .. (108.33,160.5) .. controls (108.33,164.09) and (105.42,167) .. (101.83,167) .. controls (98.24,167) and (95.33,164.09) .. (95.33,160.5) -- cycle ;
\draw   (95.33,180.5) .. controls (95.33,176.91) and (98.24,174) .. (101.83,174) .. controls (105.42,174) and (108.33,176.91) .. (108.33,180.5) .. controls (108.33,184.09) and (105.42,187) .. (101.83,187) .. controls (98.24,187) and (95.33,184.09) .. (95.33,180.5) -- cycle ;
\draw  [dash pattern={on 0.84pt off 2.51pt}]  (101.33,197) -- (101.33,217) ;
\draw   (95.33,229.5) .. controls (95.33,225.91) and (98.24,223) .. (101.83,223) .. controls (105.42,223) and (108.33,225.91) .. (108.33,229.5) .. controls (108.33,233.09) and (105.42,236) .. (101.83,236) .. controls (98.24,236) and (95.33,233.09) .. (95.33,229.5) -- cycle ;
\draw  [dash pattern={on 4.5pt off 4.5pt}] (81.33,108.5) .. controls (81.33,104.08) and (84.92,100.5) .. (89.33,100.5) -- (113.33,100.5) .. controls (117.75,100.5) and (121.33,104.08) .. (121.33,108.5) -- (121.33,239) .. controls (121.33,243.42) and (117.75,247) .. (113.33,247) -- (89.33,247) .. controls (84.92,247) and (81.33,243.42) .. (81.33,239) -- cycle ;
\draw  [dash pattern={on 0.84pt off 2.51pt}]  (71.33,170) -- (331.5,170) ;
\draw  [color={rgb, 255:red, 0; green, 92; blue, 255 }  ,draw opacity=1 ] (208.33,86.5) .. controls (208.33,82.91) and (211.24,80) .. (214.83,80) .. controls (218.42,80) and (221.33,82.91) .. (221.33,86.5) .. controls (221.33,90.09) and (218.42,93) .. (214.83,93) .. controls (211.24,93) and (208.33,90.09) .. (208.33,86.5) -- cycle ;
\draw [color={rgb, 255:red, 0; green, 92; blue, 255 }  ,draw opacity=1 ] [dash pattern={on 0.84pt off 2.51pt}]  (121.33,170) -- (211.33,90) ;
\draw  [draw opacity=0] (143.78,150.1) .. controls (148.48,155.4) and (151.33,162.37) .. (151.33,170) .. controls (151.33,170.05) and (151.33,170.09) .. (151.33,170.14) -- (121.33,170) -- cycle ; \draw  [color={rgb, 255:red, 0; green, 92; blue, 255 }  ,draw opacity=1 ] (143.78,150.1) .. controls (148.48,155.4) and (151.33,162.37) .. (151.33,170) .. controls (151.33,170.05) and (151.33,170.09) .. (151.33,170.14) ;  
\draw  [color={rgb, 255:red, 255; green, 0; blue, 0 }  ,draw opacity=1 ] (278.33,66.5) .. controls (278.33,62.91) and (281.24,60) .. (284.83,60) .. controls (288.42,60) and (291.33,62.91) .. (291.33,66.5) .. controls (291.33,70.09) and (288.42,73) .. (284.83,73) .. controls (281.24,73) and (278.33,70.09) .. (278.33,66.5) -- cycle ;
\draw  [color={rgb, 255:red, 255; green, 0; blue, 0 }  ,draw opacity=1 ] (278.33,93.5) .. controls (278.33,89.91) and (281.24,87) .. (284.83,87) .. controls (288.42,87) and (291.33,89.91) .. (291.33,93.5) .. controls (291.33,97.09) and (288.42,100) .. (284.83,100) .. controls (281.24,100) and (278.33,97.09) .. (278.33,93.5) -- cycle ;
\draw [color={rgb, 255:red, 255; green, 0; blue, 0 }  ,draw opacity=1 ] [dash pattern={on 0.84pt off 2.51pt}]  (121.33,170) -- (278.62,70.14) ;
\draw [color={rgb, 255:red, 255; green, 0; blue, 0 }  ,draw opacity=1 ] [dash pattern={on 0.84pt off 2.51pt}]  (121.33,170) -- (278.05,135) ;
\draw  [draw opacity=0] (178.04,134.1) .. controls (184.53,144.33) and (188.33,156.44) .. (188.44,169.43) -- (121.33,170) -- cycle ; \draw  [color={rgb, 255:red, 255; green, 0; blue, 0 }  ,draw opacity=1 ] (178.04,134.1) .. controls (184.53,144.33) and (188.33,156.44) .. (188.44,169.43) ;  
\draw  [draw opacity=0] (200.84,132.96) .. controls (206.09,144.21) and (209.02,156.76) .. (209.02,170) .. controls (209.02,170.05) and (209.02,170.1) .. (209.02,170.15) -- (121.33,170) -- cycle ; \draw  [color={rgb, 255:red, 255; green, 0; blue, 0 }  ,draw opacity=1 ] (200.84,132.96) .. controls (206.09,144.21) and (209.02,156.76) .. (209.02,170) .. controls (209.02,170.05) and (209.02,170.1) .. (209.02,170.15) ;  
\draw  [dash pattern={on 0.84pt off 2.51pt}]  (284.76,105.71) -- (284.76,121.57) ;
\draw  [color={rgb, 255:red, 255; green, 0; blue, 0 }  ,draw opacity=1 ] (278.33,133.5) .. controls (278.33,129.91) and (281.24,127) .. (284.83,127) .. controls (288.42,127) and (291.33,129.91) .. (291.33,133.5) .. controls (291.33,137.09) and (288.42,140) .. (284.83,140) .. controls (281.24,140) and (278.33,137.09) .. (278.33,133.5) -- cycle ;
\draw [color={rgb, 255:red, 255; green, 0; blue, 0 }  ,draw opacity=1 ] [dash pattern={on 0.84pt off 2.51pt}]  (121.33,170) -- (278.05,96.14) ;
\draw  [draw opacity=0] (251.85,140.87) .. controls (253.93,150.25) and (255.03,160) .. (255.03,170) .. controls (255.03,170.11) and (255.03,170.22) .. (255.03,170.33) -- (121.33,170) -- cycle ; \draw  [color={rgb, 255:red, 255; green, 0; blue, 0 }  ,draw opacity=1 ] (251.85,140.87) .. controls (253.93,150.25) and (255.03,160) .. (255.03,170) .. controls (255.03,170.11) and (255.03,170.22) .. (255.03,170.33) ;  
\draw  [color={rgb, 255:red, 255; green, 0; blue, 0 }  ,draw opacity=1 ][dash pattern={on 4.5pt off 4.5pt}] (264.76,56.32) .. controls (264.76,51.9) and (268.34,48.32) .. (272.76,48.32) -- (296.76,48.32) .. controls (301.18,48.32) and (304.76,51.9) .. (304.76,56.32) -- (304.76,143.86) .. controls (304.76,148.28) and (301.18,151.86) .. (296.76,151.86) -- (272.76,151.86) .. controls (268.34,151.86) and (264.76,148.28) .. (264.76,143.86) -- cycle ;
\draw  [dash pattern={on 0.84pt off 2.51pt}]  (224.43,189.29) -- (241.43,189.29) ;

\draw (86.33,79) node [anchor=north west][inner sep=0.75pt]  [font=\large] [align=left] {Bob};
\draw (196.33,59) node [anchor=north west][inner sep=0.75pt]  [font=\large] [align=left] {Alice};
\draw (266.76,28.43) node [anchor=north west][inner sep=0.75pt]  [font=\large] [align=left] {Eve};
\draw (144.33,175.57) node [anchor=north west][inner sep=0.75pt]  [font=\large,color={rgb, 255:red, 0; green, 92; blue, 255 }  ,opacity=1 ]  {$\theta $};
\draw (181.5,173.23) node [anchor=north west][inner sep=0.75pt]  [font=\large,color={rgb, 255:red, 255; green, 0; blue, 0 }  ,opacity=1 ]  {$\hat{\theta }_{0}$};
\draw (202.42,173.4) node [anchor=north west][inner sep=0.75pt]  [font=\large,color={rgb, 255:red, 255; green, 0; blue, 0 }  ,opacity=1 ]  {$\hat{\theta }_{1}$};
\draw (248.5,173.23) node [anchor=north west][inner sep=0.75pt]  [font=\large,color={rgb, 255:red, 255; green, 0; blue, 0 }  ,opacity=1 ]  {$\hat{\theta }_{L-1}$};

\end{tikzpicture}
\end{center}
\caption{System model considered, in which Alice is equipped with a single antenna while both Bob and Eve are equipped with an array of antennas.}
\label{fig:systemmodel}
\end{figure}

Let us consider the system model depicted in Fig. \ref{fig:systemmodel}. Alice is equipped with a single transmitting antenna, while the receiver Bob is equipped with a digital \ac{ULA} of receiving antennas, formed by $M$ elements uniformly spaced by a distance $d$. We assume the far-field condition holds, i.e., $B\ll f_{c}$, where $B$ and $f_{c}$ are the bandwidth and the carrier frequency, respectively, and $s(t)= Re\{s_{0}(t)e^{j2\pi f_{c}t}\}$ is the narrowband source signal. Then, the time delay of the arrival at the $m$-th
element is simply $\Delta t_{m}=\frac{md}{c}\sin\theta$, where $c=\lambda f_{c}$
is the velocity of propagation, $\lambda$ is the wavelength, and $\theta$ is the angle of arrival (AoA) to be estimated.

At the receiver side, the received baseband signal at the $m$-th
element is given by

\begin{equation}
x_{m}(t)=s_{0}(t-\Delta t_{m})e^{-j2\pi f_{c}\Delta t_{m}}+n(t),m\in\{0,\ldots,M-1\},
\end{equation}
whose \textit{discrete form} can be approximated as 
\begin{eqnarray}
x_{m}[i] & \simeq & s_{0}[i]e^{-j\frac{2\pi}{\lambda}md\sin\theta}+n[i]\\
 & = & s_{0}[i]a_{m}(\theta)+n[i],m\in\{0,\ldots,M-1\}\label{eq:xm}
\end{eqnarray}
where $a_{m}(\theta)=e^{-j\frac{2\pi}{\lambda}md\sin\theta}$. Let
us define $\kappa=\frac{2\pi}{\lambda}d$ and rewrite \eqref{eq:xm}
in vectorial form, that is, 
\begin{equation}
\mathbf{x}[i]=\mathbf{a}s_{0}[i]+\mathbf{n}[i],
\end{equation}
where 
\begin{equation}
\mathbf{a}=[1\quad e^{-j\kappa\sin(\theta)}\quad e^{-j\kappa2\sin(\theta)}\cdots e^{-j\kappa(M-1)\sin(\theta)}]^{T}
\end{equation}
is the \textit{steering vector} and $\mathbf{n}$ is a Gaussian circularly symmetric noise vector.
In the following, to lighten the notation, we will no longer write the dependence on time instant $i$, assuming it implicitly.

Various methods can be employed to estimate the AoA from $\mathbf{x}$, given $s_{0}$. Examples include the delay-and-sum method, minimum variance distortionless response (MVDR), and the multiple signal classification (MUSIC) algorithm \cite{Godara:DS:1997,Capon:MDVR:1969,Schmidt:MUSIC:1986}. Notably, the widely used MUSIC method leverages the noise subspace for estimation and is regarded as a high-resolution technique. Due to these advantages, we focus on the MUSIC method in this paper. Note that we consider 2D systems in the considered scenarios, an extension to 3D systems will be discussed in Section \ref{sec:num}.

\section{Resistance to impersonation attacks Based on AoA}

\label{sec:AoA}

In this section, we prove the following proposition for an attacker with single antenna, two antennas and then generalize to the general case.

\begin{prop}
An adversary cannot impersonate the AoA of the legitimate transmitter as long as their angles are not identical.
\label{prop:AoA:resistance:gen} 
\end{prop}


\subsection{Single-antenna attacker}
\label{subsec:impersonation}

Let us suppose a network of static nodes for which Bob records the AoA-based signature as part of the authentication process. An adversary
with an angle $\hat{\theta}$ and associated steering vector $\hat{\mathbf{a}}=[1\quad e^{-j\kappa\sin(\hat{\theta})}\quad e^{-j\kappa2\sin(\hat{\theta})}\cdots e^{-j\kappa(M-1)\sin(\hat{\theta})}]^{T}$
can try to perform a Sybil attack. We prove in the following that impersonation is impossible if $\hat{\theta}\neq\theta$.

For this case, we will prove an adversary with a single antenna cannot impersonate the AoA of the
legitimate transmitter as long as their angles are not identical.

\begin{proof} At any time instant $i$, the signal received by Bob
from the legitimate transmitter can be expressed as 
\begin{equation}
\mathbf{x}=\mathbf{a}s_{0}+\mathbf{n}.
\end{equation}

A single-antenna adversary with true angle $\hat{\theta}$ and associated steering vector $\hat{\mathbf{a}}$ can precode its signal by introducing some complex precoding factor $q$ to try to impersonate the legitimate user, so that the legitimate receiver sees the signal expressed below
\begin{equation}
\hat{\mathbf{x}}=\hat{\mathbf{a}}qs_{0}+\hat{\mathbf{n}}.\label{eq: att}
\end{equation}

The mean square error (MSE) between the signals received from the
legitimate and adversarial transmitters is thus given by 
\begin{align}
\zeta & =\mathbb{E}(||\mathbf{x}-\hat{\mathbf{x}}||^{2})\nonumber \\
 & =\mathbb{E}\left(|s_{0}|^{2}\left(\mathbf{a}^{H}\mathbf{a}-\mathbf{a}^{H}q\hat{\mathbf{a}}-\hat{\mathbf{a}}^{H}q^{*}\mathbf{a}+\hat{\mathbf{a}}^{H}q^{*}q\hat{\mathbf{a}}\right)+\right.\nonumber \\
 & \mathrel{\phantom{=}}\left.||\mathbf{n}||^{2}+||\hat{\mathbf{n}}||^{2}\right).
\end{align}

Let us denote by $\delta_{n}$ and $\delta_{\hat{n}}$ the SNR of the legitimate and adversarial transmitters. The above equation then
becomes 
\begin{equation}
\zeta=|s_{0}|^{2}\left(\mathbf{a}^{H}\mathbf{a}-q\mathbf{a}^{H}\hat{\mathbf{a}}-q^{*}\hat{\mathbf{a}}^{H}\mathbf{a}+q^{*}q\hat{\mathbf{a}}^{H}\hat{\mathbf{a}}+\frac{1}{\delta_{n}}+\frac{1}{\delta_{\hat{n}}}\right).
\end{equation}

Without loss of generality, we can assume a unitary power pilot signal
and $q$ corresponding to a general complex precoding factor, i.e., $q=\beta e^{j\phi}$.
We then obtain 
\begin{align}
\zeta & =\mathbf{a}^{H}\mathbf{a}-q\mathbf{a}^{H}\hat{\mathbf{a}}-q^{*}\hat{\mathbf{a}}^{H}\mathbf{a}+\hat{\mathbf{a}}^{H}\hat{\mathbf{a}}+\frac{1}{\delta_{n}}+\frac{1}{\delta_{\hat{n}}}\nonumber \\
 & =\mathbf{a}^{H}\mathbf{a}+\beta^{2}\hat{\mathbf{a}}^{H}\hat{\mathbf{a}}-\beta e^{j\phi}\mathbf{a}^{H}\hat{\mathbf{a}}-\beta e^{-j\phi}\hat{\mathbf{a}}^{H}\mathbf{a}+\frac{1}{\delta_{n}}+\frac{1}{\delta_{\hat{n}}}.\label{eq:zeta}
\end{align}
By the definition of the steering vectors, we get 
\begin{align}
\mathbf{a}^{H}\mathbf{a} & =[1\quad \cdots e^{j\kappa(M-1)\sin(\theta)}]\left[\begin{array}{c}
1\\
\vdots\\
e^{-j\kappa(M-1)\sin(\theta)}
\end{array}\right]\nonumber \\
 & =M,\label{eq:compute:1terms}
\end{align}
and 
\begin{align}
\mathbf{a}^{H}\hat{\mathbf{a}} & =[1\quad \cdots e^{j\kappa(M-1)\sin(\theta)}]\left[\begin{array}{c}
1\\
\vdots\\
e^{-j\kappa(M-1)\sin(\hat{\theta})}
\end{array}\right]\\
 & =1+e^{j\kappa(\sin(\theta)-\sin(\hat{\theta}))}+\ldots+e^{j\kappa(M-1)(\sin(\theta)-\sin(\hat{\theta}))}\nonumber \\
 & =1+e^{j\kappa\alpha}+\ldots+e^{j\kappa(M-1)\alpha},
\end{align}
where $\alpha=\sin(\theta)-\sin(\hat{\theta})$. Similarly, we obtain
\begin{equation}
\hat{\mathbf{a}}^{H}\hat{\mathbf{a}}=M,
\end{equation}
and 
\begin{equation}
\hat{\mathbf{a}}^{H}\mathbf{a}=1+e^{-j\kappa\alpha}+\ldots+e^{-j\kappa(M-1)\alpha}.\label{eq:compute:4terms}
\end{equation}
Substituting \eqref{eq:compute:1terms}-\eqref{eq:compute:4terms}
into \eqref{eq:zeta} yields 
\begin{align}
\zeta & =(\beta^{2}+1)M-\beta\left(e^{j\phi}+e^{-j\phi}\right)-\ldots\nonumber \\
 & \mathrel{\phantom{=}}-\beta\left(e^{j(\kappa(M-1)\alpha+\phi)}+e^{-j(\kappa(M-1)\alpha+\phi)}\right)+\frac{1}{\delta_{n}}+\frac{1}{\delta_{\hat{n}}}.
\end{align}

From the properties of the complex exponential function, we can obtain
\begin{align}
\zeta & =\underbrace{(\beta^{2}+1)M-2\beta\left(\cos(\phi)+\ldots+\cos(\kappa(M-1)\alpha+\phi)\right)}_{\Delta}\label{eq:before:trans}\nonumber\\
 & +\frac{1}{\delta_{n}}+\frac{1}{\delta_{\hat{n}}},
\end{align}
and define $\Delta$ as in \eqref{eq:before:trans}.
We can then consider two cases:

\textbf{Case 1} $\alpha=0$: The angle of the adversary is the same
as that of the legitimate user. We have 
\begin{equation}
\cos(\phi)+\ldots+\cos(\kappa(M-1)\alpha+\phi)=M\cos(\phi),
\end{equation}
and thus 
\begin{align}
\Delta & =(\beta^{2}+1)M-2\beta M\cos(\phi)=M(\beta^{2}+1-2\beta\cos(\phi))\\
 & =M((\beta-\cos(\phi))^{2}+\sin^{2}(\phi))\geq0.
\end{align}

We can easily see that $\zeta$ achieves the minimum at $\zeta=\frac{1}{\delta_{n}}+\frac{1}{\delta_{\hat{n}}}$ if and only if $\Delta = 0$, that is, for $\phi=0$ and $\beta=1$, or simply $q=1$.

\textbf{Case 2} $\alpha\neq0$: The angle of the adversary is different
from that of the legitimate one. Based on the result of the sum of
cosine \cite{Samuel:Funfact:1986}, we have 
\begin{align}
&\cos(\phi)+\ldots+\cos(\kappa(M-1)\alpha+\phi)\nonumber\\&=\frac{\sin(\frac{M\kappa\alpha}{2})}{\sin(\frac{\kappa\alpha}{2})}\cos((M-1)\kappa\frac{\alpha}{2}+\phi),
\end{align}
and thus 
\begin{equation}
\Delta=(\beta^{2}+1)M-2\beta\frac{\sin(\frac{M\kappa\alpha}{2})}{\sin(\frac{\kappa\alpha}{2})}\cos((M-1)\kappa\frac{\alpha}{2}+\phi).
\end{equation}

Consider the following partial derivatives

\begin{equation}
\frac{\partial\zeta}{\partial\beta}=2\beta M-2\frac{\sin(\frac{M\kappa\alpha}{2})}{\sin(\frac{\kappa\alpha}{2})}\cos\left((M-1)\frac{\kappa\alpha}{2}+\phi\right),
\end{equation}

\begin{equation}
\frac{\partial\zeta}{\partial\phi}=2\beta\frac{\sin(\frac{M\kappa\alpha}{2})}{\sin(\frac{\kappa\alpha}{2})}\sin\left((M-1)\frac{\kappa\alpha}{2}+\phi\right),
\end{equation}
we achieve the critical points at $\frac{\partial\zeta}{\partial\beta}=0$
and $\frac{\partial\zeta}{\partial\phi}=0$, therefore

\begin{equation}
\beta=\frac{1}{M}\frac{\sin(\frac{M\kappa\alpha}{2})}{\sin(\frac{\kappa\alpha}{2})}\cos\left((M-1)\frac{\kappa\alpha}{2}+\phi\right),
\end{equation}
and 
\begin{equation}
\phi=-(M-1)\frac{\kappa\alpha}{2}+u\pi,
\end{equation}
 where $u$ is an integer.

The second derivatives are as follows

\begin{equation}
\frac{\partial^{2}\zeta}{\partial\beta^{2}}=2M,
\end{equation}

\begin{equation}
\frac{\partial^{2}\zeta}{\partial\phi^{2}}=2\beta\frac{\sin(\frac{M\kappa\alpha}{2})}{\sin(\frac{\kappa\alpha}{2})}\cos\left((M-1)\frac{\kappa\alpha}{2}+\phi\right),
\end{equation}

\begin{equation}
\frac{\partial^{2}\zeta}{\partial\beta\partial\phi}=\frac{\partial^{2}\zeta}{\partial\phi\partial\beta}=2\frac{\sin(\frac{M\kappa\alpha}{2})}{\sin(\frac{\kappa\alpha}{2})}\sin\left((M-1)\frac{\kappa\alpha}{2}+\phi\right).
\end{equation}

We can obtain

\begin{align*}
D & =\left|\begin{array}{cc}
\frac{\partial^{2}\zeta}{\partial\beta^{2}} & \frac{\partial^{2}\zeta}{\partial\beta\partial\phi}\\
\frac{\partial^{2}\zeta}{\partial\phi\partial\beta} & \frac{\partial^{2}\zeta}{\partial\phi^{2}}
\end{array}\right|\\
 & =4M\beta\frac{\sin(\frac{M\kappa\alpha}{2})}{\sin(\frac{\kappa\alpha}{2})}\cos\left((M-1)\frac{\kappa\alpha}{2}+\phi\right)\\
 & -4\frac{\sin^{2}(\frac{M\kappa\alpha}{2})}{\sin^{2}(\frac{\kappa\alpha}{2})}\sin^{2}\left((M-1)\frac{\kappa\alpha}{2}+\phi\right).
\end{align*}

If $\phi=-(M-1)\frac{\kappa\alpha}{2}+u\pi$ where $u$ is an even
number thus $\beta=\frac{1}{M}\frac{\sin(\frac{M\kappa\alpha}{2})}{\sin(\frac{\kappa\alpha}{2})}$
and 
\begin{equation}
D =4M\beta\frac{\sin(\frac{M\kappa\alpha}{2})}{\sin(\frac{\kappa\alpha}{2})} =4\frac{\sin^{2}(\frac{M\kappa\alpha}{2})}{\sin^{2}(\frac{\kappa\alpha}{2})}.
\end{equation}

Note that $\alpha\neq0$, thus $D>0$. Moreover $\frac{\partial^{2}\zeta}{\partial\beta^{2}}=2M>0$,
thus we achieve the local minimum.

Similarly, if $\phi=-(M-1)\frac{\kappa\alpha}{2}+u\pi$ where $u$
is an odd number, thus $\beta=-\frac{1}{M}\frac{\sin(\frac{M\kappa\alpha}{2})}{\sin(\frac{\kappa\alpha}{2})}$
and 
\begin{equation}
D =-4M\beta\frac{\sin(\frac{M\kappa\alpha}{2})}{\sin(\frac{\kappa\alpha}{2})} =4\frac{\sin^{2}(\frac{M\kappa\alpha}{2})}{\sin^{2}(\frac{\kappa\alpha}{2})}.
\end{equation}

We also notice that $\alpha\neq0$, therefore $D>0$. Since $\frac{\partial^{2}\zeta}{\partial\beta^{2}}=2M>0$,
we can also achieve the local minimum. It is worth noting that these local minima are bounded away from zero as shown in \eqref{eq:before:trans}. 

\end{proof}

\subsection{Two Single-antenna
Attackers}

In the case in which Alice has one antenna and Eve has two
distributed antennas - corresponding to two sources, we will also prove that an adversary with two antennas cannot impersonate the AoA of the
legitimate transmitter unless their angles are identical.

Bob receives a signal from Alice as
\begin{equation}
\mathbf{x}=\mathbf{a}s+\mathbf{n},
\end{equation}

where $\mathbf{a}=\left[\begin{array}{c}
1\\
e^{-j\kappa\sin(\theta)}\\
\vdots\\
e^{-j(M-1)\kappa\sin(\theta)}
\end{array}\right].$

The adversary has two sources and thus 
\begin{equation}
\hat{\mathbf{x}}=\hat{\mathbf{A}}\mathbf{Q}\hat{\mathbf{s}}+\hat{\mathbf{n}}.
\end{equation}

Note that $\mathbf{Q}\hat{\mathbf{s}}$ is a vector, without loss of generality, we can assume that the pilot signal is known and can thus equivalently write the equation as the following
\begin{equation}
\hat{\mathbf{x}}=\hat{\mathbf{A}}\mathbf{\hat{q}}s+\hat{\mathbf{n}},
\end{equation}
where $\mathbf{\hat{q}}=[\hat{q}_{0}\,\hat{q}_{1}]^{T}$.

The MSE is thus given by

\begin{align}
\nonumber
\zeta & =\mathbb{E}(||\mathbf{x}-\hat{\mathbf{x}}||^{2})\\
\nonumber
 & =\mathbb{E}(s^{*}(\mathbf{a}-\hat{\mathbf{A}}\mathbf{\mathbf{\hat{q}}})^{H}({\mathbf{a}}-\hat{\mathbf{A}}\mathbf{\mathbf{\hat{q}}})s+||\mathbf{n}||^{2}+||\hat{\mathbf{n}}||^{2})\\
 \nonumber
 & =\mathbb{E}((\mathbf{a}^{H}\mathbf{a}-\mathbf{a}^{H}\mathbf{\hat{\mathbf{A}}}\mathbf{\mathbf{\hat{q}}}-\mathbf{\mathbf{\hat{q}}}^{H}\hat{\mathbf{A}}^{H}\mathbf{a}+\mathbf{\mathbf{\hat{q}}}^{H}\hat{\mathbf{A}}^{H}\mathbf{\hat{\mathbf{A}}}\mathbf{\mathbf{\hat{q}}})|\mathbf{s}|^{2}\\&+||\mathbf{n}||^{2}+||\hat{\mathbf{n}}||^{2})\\
 & =\mathbf{a}^{H}\mathbf{a}-\mathbf{a}^{H}\mathbf{\hat{\mathbf{A}}}\mathbf{\mathbf{\hat{q}}}-\mathbf{\mathbf{\hat{q}}}^{H}\hat{\mathbf{A}}^{H}\mathbf{a}+\mathbf{\mathbf{\hat{q}}}^{H}\hat{\mathbf{A}}^{H}\mathbf{\hat{\mathbf{A}}}\mathbf{\mathbf{\hat{q}}+}\frac{1}{\delta_{n}}+\frac{1}{\delta_{\hat{n}}}.
 \label{eq:MSE}
\end{align}

Similarly, we can prove that 
\begin{equation}
\zeta=M\left((\beta_{0}+\beta_{1}-\cos(\phi_{0}))^{2}+\sin^{2}(\phi_{0})\right)+\frac{1}{\delta_{n}}+\frac{1}{\delta_{\hat{n}}}
\end{equation}
only achieve the global minimum at $\hat{q}_{0}+\hat{q}_{1}=1$ and $\hat{q}_{0}, \hat{q}_{1}$ are real.
For the detailed proof, interested can refer to Appendix \ref{app:2ants}.

\subsection{Multiple
Attackers}
Assume that Alice has a single antenna and there are $L$ signal sources at Eve, which are considered as multiple distributed antennas.
Assume the pilot signal is known, thus we can obtain
\begin{equation}
\hat{\mathbf{x}}=\hat{\mathbf{A}}\mathbf{\hat{q}}s+\hat{\mathbf{n}},
\end{equation}
where $\mathbf{\hat{q}}=[\hat{q}_{0}\,\hat{q}_{1}, \cdots, \hat{q}_{L-1}]^{T}$

The MSE is thus given by

\begin{align}
\zeta & =\mathbb{E}(||\mathbf{x}-\hat{\mathbf{x}}||^{2})\\
 & =\mathbb{E}(s^{*}(\mathbf{a}-\hat{\mathbf{A}}\mathbf{\mathbf{\hat{q}}})^{H}(\mathbf{A}-\hat{\mathbf{A}}\mathbf{\mathbf{\hat{q}}})s+||\mathbf{n}||^{2}+||\hat{\mathbf{n}}||^{2})\\
 & =\mathbb{E}((\mathbf{a}^{H}\mathbf{a}-\mathbf{a}^{H}\mathbf{\hat{\mathbf{A}}}\mathbf{\mathbf{\hat{q}}}-\mathbf{\mathbf{\hat{q}}}^{H}\hat{\mathbf{A}}^{H}\mathbf{a}+\mathbf{\mathbf{\hat{q}}}^{H}\hat{\mathbf{A}}^{H}\mathbf{\hat{\mathbf{A}}}\mathbf{\mathbf{\hat{q}}})|\mathbf{s}|^{2}\nonumber\\&+||\mathbf{n}||^{2}+||\hat{\mathbf{n}}||^{2})\\
 & =\mathbf{a}^{H}\mathbf{a}-\mathbf{a}^{H}\mathbf{\hat{\mathbf{A}}}\mathbf{\mathbf{\hat{q}}}-\mathbf{\mathbf{\hat{q}}}^{H}\hat{\mathbf{A}}^{H}\mathbf{a}+\mathbf{\mathbf{\hat{q}}}^{H}\hat{\mathbf{A}}^{H}\mathbf{\hat{\mathbf{A}}}\mathbf{\mathbf{\hat{q}}+}\frac{1}{\delta_{n}}+\frac{1}{\delta_{\hat{n}}}.
\end{align}

From the definition, we achieve
\begin{equation}
\mathbf{a}^{H}\mathbf{a}=M,
\end{equation}

\begin{align}
&\mathbf{a}^{H}\mathbf{\hat{\mathbf{A}}}\mathbf{\mathbf{\hat{q}}=}\nonumber\\ & (1+e^{j\kappa(\sin(\theta)-\sin(\hat{\theta}_0))}+\ldots+e^{j(M-1)\kappa(\sin(\theta)-\sin(\hat{\theta}_0))})\hat{q}_{0}\nonumber\\&+(1+e^{j\kappa(\sin(\theta)-\sin(\hat{\theta}_{1}))}+\ldots+e^{j(M-1)\kappa(\sin(\theta)-\sin(\hat{\theta}_{1}))})\hat{q}_{1}\nonumber\\&+\ldots+(1+\ldots+e^{j(M-1)\kappa(\sin(\theta)-\sin(\hat{\theta}_{1}))})\hat{q}_{L-1},
\end{align}


\begin{align}
\mathbf{\mathbf{\hat{q}}}^{H}\hat{\mathbf{A}}^{H}\mathbf{a}= (1+\ldots+e^{-j(M-1)\kappa(\sin(\theta)-\sin(\hat{\theta}_{1}))})\hat{q}_{1}^{\ast}\nonumber \\+\ldots+(1+\ldots+e^{-j(M-1)\kappa(\sin(\theta)-\sin(\hat{\theta}_{1}))})\hat{q}_{L-1}^{\ast},
\end{align}

\begin{equation}
\mathbf{\mathbf{\hat{q}}}^{H}\hat{\mathbf{A}}^{H}\mathbf{\hat{\mathbf{A}}}\mathbf{\mathbf{\hat{q}}}=\mathbf{\mathbf{\hat{q}}}^{H}\mathbf{G}\mathbf{\mathbf{\hat{q}}},
\end{equation}
where $\mathbf{G}$ is defined as 
\begin{equation}
\mathbf{G}=\left[\begin{array}{cccc}
g_{11} & g_{12} & \ldots & g_{1L}\\
g_{21} & g_{22} & \ldots & g_{2L}\\
\vdots & \vdots & \ddots & \vdots\\
g_{L1} & g_{L2} & \ldots & g_{LL}
\end{array}\right],
\end{equation}
where $g_{ll}=M$ and $g_{lz}=1+e^{-j\kappa(\sin\hat{\theta}_{l}-\sin\hat{\theta}_{z})}+\ldots+e^{-j(M-1)\kappa(\sin\hat{\theta}_{l}-\sin\hat{\theta}_{z})}$.
\begin{align}
\mathbf{\mathbf{\hat{q}}}^{H}\hat{\mathbf{A}}^{H}\mathbf{\hat{\mathbf{A}}}\mathbf{\mathbf{\hat{q}}}&= g_{11}\hat{q}_{0}^{2}+g_{22}\hat{q}_{1}^{2}+\ldots+g_{LL}\hat{q}_{L-1}^{2}\nonumber\\&+(g_{12}\hat{q}_{1}+g_{13}\hat{q}_{2}+\ldots+g_{1L}\hat{q}_{L-1})\hat{q}_{0}^{\ast}+\ldots\nonumber\\&+(g_{L1}\hat{q}_{0}+g_{L2}\hat{q}_{1}+\ldots+g_{L(L-1)}\hat{q}_{L-2})\hat{q}_{L-1}^{\ast}\\
&= (\hat{q}_{0}^{2}+\hat{q}_{1}^{2}+\ldots+\hat{q}_{L-1}^{2})M\nonumber\\&+(g_{12}\hat{q}_{1}+g_{13}\hat{q}_{2}+\ldots+g_{1L}\hat{q}_{L-1})\hat{q}_{0}^{\ast}+\ldots\nonumber\\&+(g_{L1}\hat{q}_{0}+g_{L2}\hat{q}_{1}+\ldots+g_{L(L-1)}\hat{q}_{L-2})\hat{q}_{L-1}^{\ast}.
\end{align}

Assume the aforementioned condition hold, i.e., $\hat{\theta}_{0}=\hat{\theta}_{1}=\hat{\theta}_{L-1}=\ldots=\theta$
so that $\zeta$ can achieve the global minimum, as the cases $L=1,L=2$.
We prove in the following the latter is also true. 

Assume 
\begin{equation}
\hat{q}_{0}+\hat{q}_{1}+\ldots+\hat{q}_{L-1}=u+jv.
\end{equation}

Since $\hat{\theta}_{0}=\hat{\theta}_{1}=\hat{\theta}_{L}=\ldots=\theta$, replacing this condition to the aforementioned terms and equation, the finding of minimization boils down to the finding of the minimum of the following

\begin{align}
\zeta & =M(1-(u+jv)-(u-jv)+(u^{2}+v^{2}))\nonumber\\&+(\frac{1}{\delta_{n}}+\frac{1}{\delta_{\hat{n}}})\\
 & =M((1-u)^{2}+v^{2})+\left(\frac{1}{\delta_{n}}+\frac{1}{\delta_{\hat{n}}}\right)\geq\left(\frac{1}{\delta_{n}}+\frac{1}{\delta_{\hat{n}}}\right).
\end{align}
\\

We can easily see that $\zeta\geq(\frac{1}{\delta_{n}}+\frac{1}{\delta_{\hat{n}}})$
and only achieve the minimum at $u=1$ and $v=0$, which requires
$\hat{q}_{0}+\hat{q}_{1}+\ldots+\hat{q}_{L-1}=1$ and $\hat{\theta}_{0}=\hat{\theta}_{1}=\hat{\theta}_{L-1}=\ldots=\theta$ as in the other
cases. 

By induction, we can conclude that the proposition holds true and the impersonation attack to AoA authentication can only happen in stringent conditions as proved in the proceeding subsections. 

\section{Numerical Results and Discussion}
\label{sec:num}
In this section, we report some numerical results that confirm the analytical derivations and results.

\subsection{AoA estimation through MUSIC}

\begin{figure}[!ht]
\begin{centering}
   \includegraphics[width=0.45\textwidth]{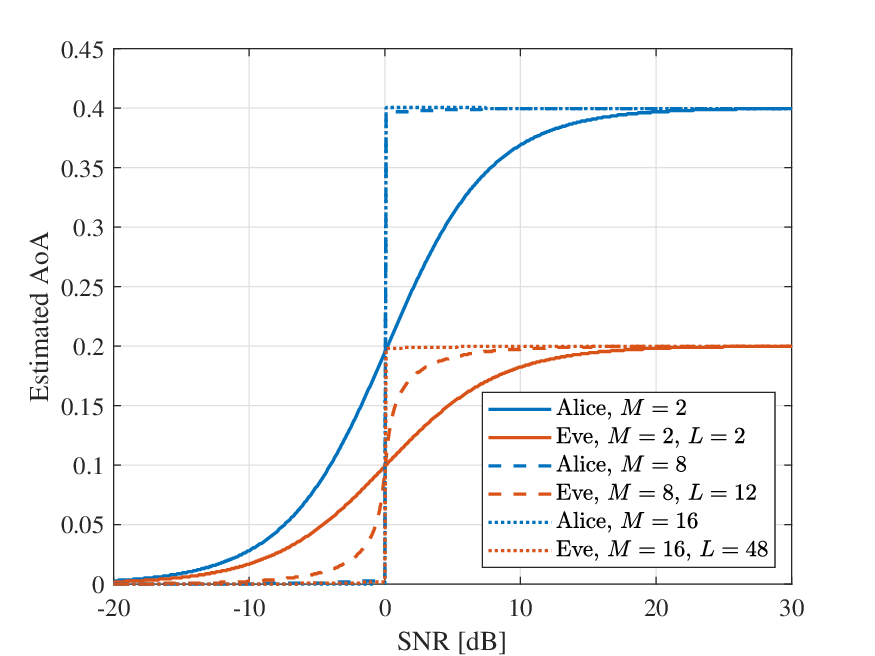}
\caption{Estimated AoA using MUSIC algorithm with 2,000 samples. Alice's AoA $\theta = 0.4$ \SI{}{rad}, AoAs of Eve's antennas $\hat{\theta}_0 = 0.2$ \SI{}{rad}, $\hat{\theta}_i = 0.2$ \SI{}{rad} and phases of Eve's precoding factors $\phi_i = 0$ $\forall i \in [0, L-1]$, sum of amplitudes of Eve's precoding factors $\sum_{i=0}^{L-1}\beta_i = 1$. }
\label{fig:MUSICest}
\par\end{centering}
\end{figure}

Initially, the limitations of the MUSIC algorithm in correctly estimating the AoA of different users are tested.
Fig. \ref{fig:MUSICest} shows the AoA estimated by MUSIC when Eve carries out an impersonation attack, considering a true AoA $\theta$ corresponding to 0.4 \SI{}{rad} for Alice and to 0.2 \SI{}{rad} for Eve, further assuming that all of the opponent's antennas are aligned along the same direction relative to the receiver. We see that MUSIC performance is highly dependent on the SNR and on the number of antennas available at the receiver. Nevertheless, it can be observed that, even when Bob is equipped with only two antennas, the estimated angles are similar only for low SNR values (less than 0 dB), due to the fact that the algorithm is not able to make an accurate estimate.

\subsection{Special case}

\begin{figure}[!ht]
\begin{centering}
   \includegraphics[width=0.45\textwidth]{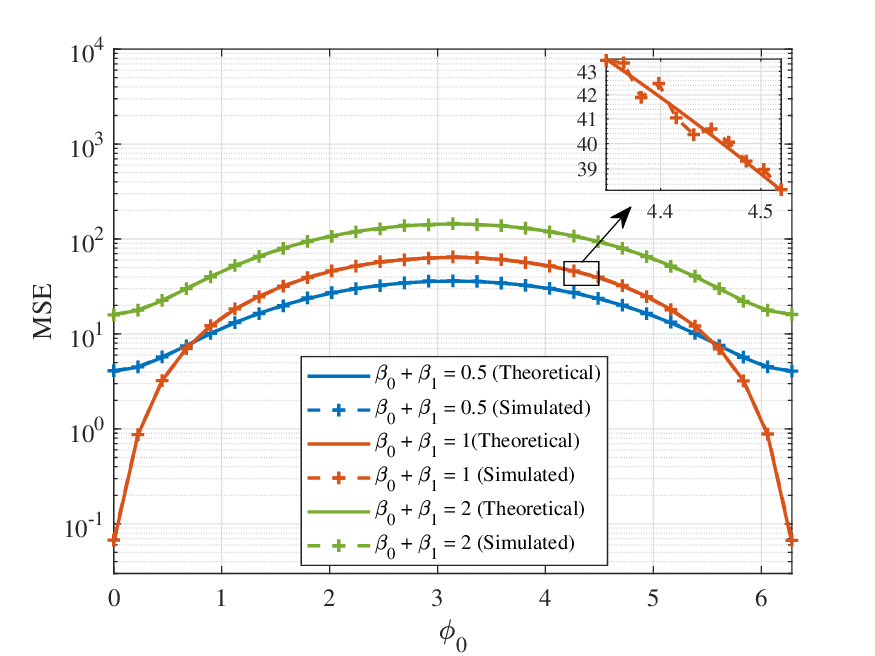}
\caption{MSE under special case scenario where AoAs of Eve's antennas are identical $\hat{\theta}_0 = \hat{\theta}_1 = 0.4$ \SI{}{rad}, the phases of Eve's precoding factors $\phi_1 = \phi_0$, Alice's AoA $\theta = 0.4$ \SI{}{rad}, SNR = 15 dB. }
\label{fig:diff_beta}
\par\end{centering}
\end{figure}

Let us first consider the special case, where Eve is equipped with two distributed antennas which have identical AoAs. Fig. \ref{fig:diff_beta} shows the MSE behavior as a function of the parameter $\phi_0$. Bob has 16 receiving antennas, and the system operates at a frequency of 2.18 GHz. We also suppose that the SNR of both Alice-Bob and Eve-Bob channels is equal to 15 dB. In this figure, the extreme scenario of Eve being in the same direction of Alice is examined. $\beta_0$ and $\beta_1$ have been computed according to previous calculations, while $\phi_1 = \phi_0$. We can observe that a minimum in the MSE is achieved only when $\phi_0$ is equal to 0 (or to $2\pi$) and the sum of $\beta_0$ and $\beta_1$ is equal to 1, as expected. Note that a null MSE is not achievable unless we consider an infinite SNR, as also evident from Eq. \eqref{eq:MSE}.
Moreover, this figure conveys another important message, proving the substantial equivalence between the theoretical curves and the simulated ones. Theoretical curves are derived using \eqref{eq:MSE}, whereas simulated curves are generated through a Monte Carlo simulation performed over 10,000 instances. For this reason, in the following figures, we will present only the results generated by the simulations. 


\begin{figure}[!ht]
\begin{centering}
 \subfigure[Eve's antennas have the same AoA]
   {\includegraphics[width=0.45\textwidth]{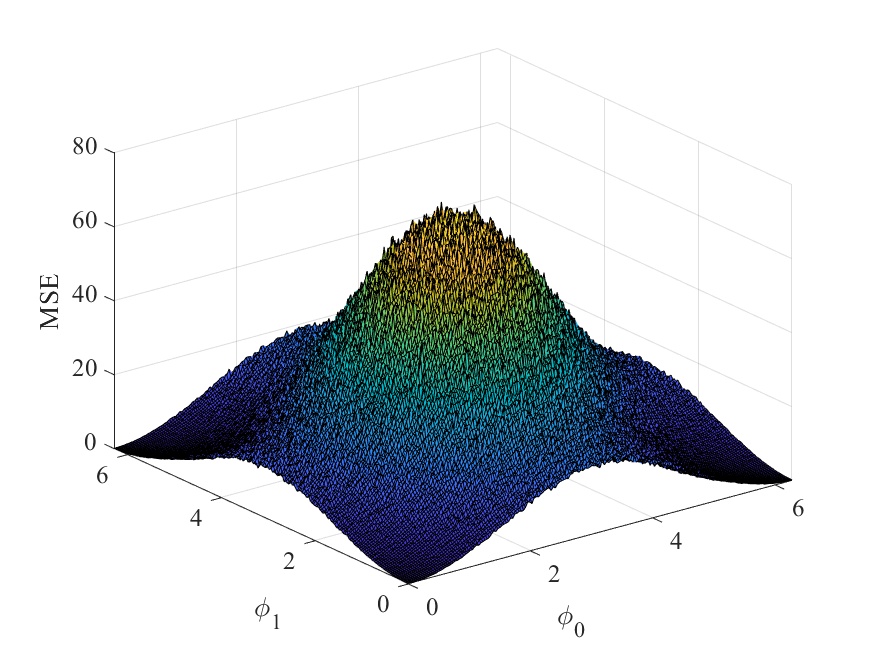}
   \label{fig:3d_same}}
 \subfigure[Eve's antennas have different AoAs]
   {\includegraphics[width=0.45\textwidth]{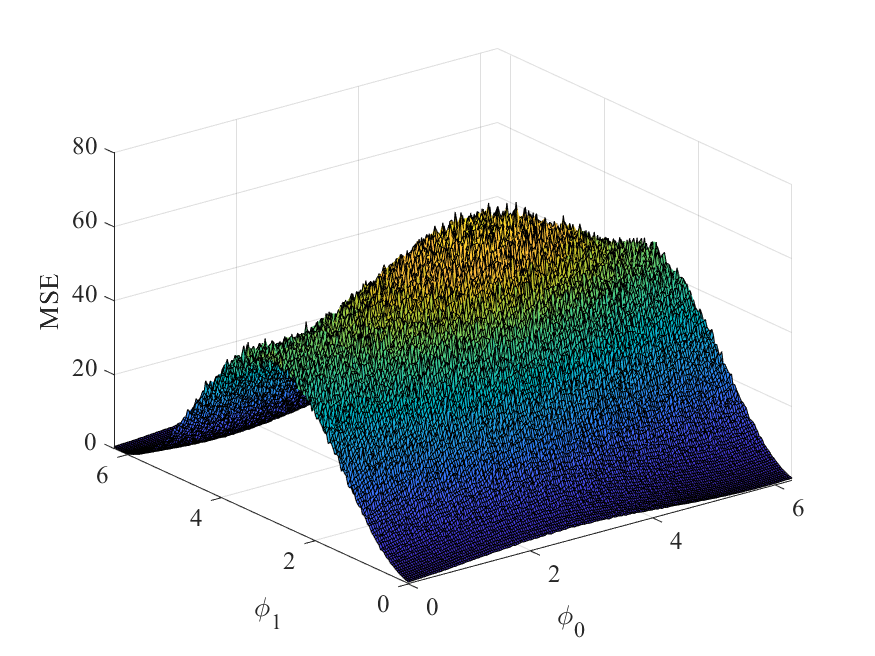}
   \label{fig:3d_diff}}
\caption{MSE under special case scenario, with SNR = 15 dB. Alice's AoA $\theta = 0.4$ \SI{}{rad}. a) Eve's antennas have the same AoA, $\hat{\theta}_0 = \hat{\theta}_1 = 0.4$ \SI{}{rad}. b)  Eve's antennas have different AoAs, $\hat{\theta}_0 = 0.39$ \SI{}{rad}, $\hat{\theta}_1 = 0.41$ \SI{}{rad}. }
\par\end{centering}
\end{figure}

Fig. \ref{fig:3d_same} and \ref{fig:3d_diff} show a comparison of what happens at the MSE when Eve's antennas have the same AoA (Fig. \ref{fig:3d_same}) or in slightly different AoAs, i.e., $\phi_0 \neq \phi_1$ (Fig. \ref{fig:3d_diff}). In both cases the sum of $\beta_0 + \beta_1 = 1$. We observe that, as expected, Figure \ref{fig:3d_diff} has no symmetry, unlikely Fig. \ref{fig:3d_same}.
In any case, these figures confirm that in both scenarios a minimum MSE is achieved when $\phi_0$ and $\phi_1$ are close to 0. 

\subsection{General case}

Let us now consider the case where Eve can have any number of antennas. This case is of particular interest since, when Eve has more antennas than Bob, she could exploit the additional spatial diversity to improve her AoA manipulation capabilities. The extra antennas provide her with more paths, potentially reducing the MSE between her and Alice's estimated AoA.

\begin{figure}[!ht]
\begin{centering}
   \includegraphics[width=0.45\textwidth]{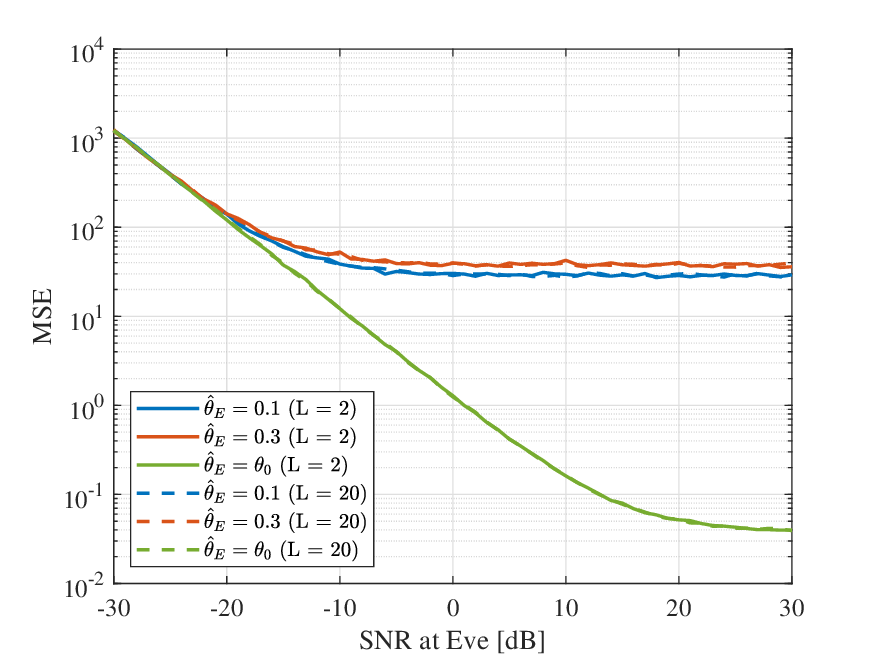}
\caption{MSE vs SNR at Eve, with SNR at Alice equal to 15 dB, number of Bob's antennas $M = 16$, Alice's AoA $\theta = 0.4$ \SI{}{rad}.}
\label{fig:MSE_vs_SNR}
\par\end{centering}
\end{figure}

Fig. \ref{fig:MSE_vs_SNR} shows the impact of the SNR at Eve on the MSE. SNR at Alice is equal to 15 dB, and we consider the most favorable situation for the attacker, i.e., $\phi_i = 0 \forall i \in[0, L-1], \sum_{i=0}^{L-1} \beta_i = 1$. We suppose that all Eve's antennas have the same AoA denoted as $\hat{\theta}_E$, i.e., $\hat{\theta_0} = \cdots = \hat{\theta}_{L-1} = {\hat{\theta}_E}$. As evident from \eqref{eq:mse}, a higher SNR leads to a lower MSE. In this figure, we can also have a first evaluation of the role of Eve's number of antennas ($L$), which, however, appears to be negligible.


\begin{figure}[!ht]
\begin{centering}
   \includegraphics[width=0.45\textwidth]{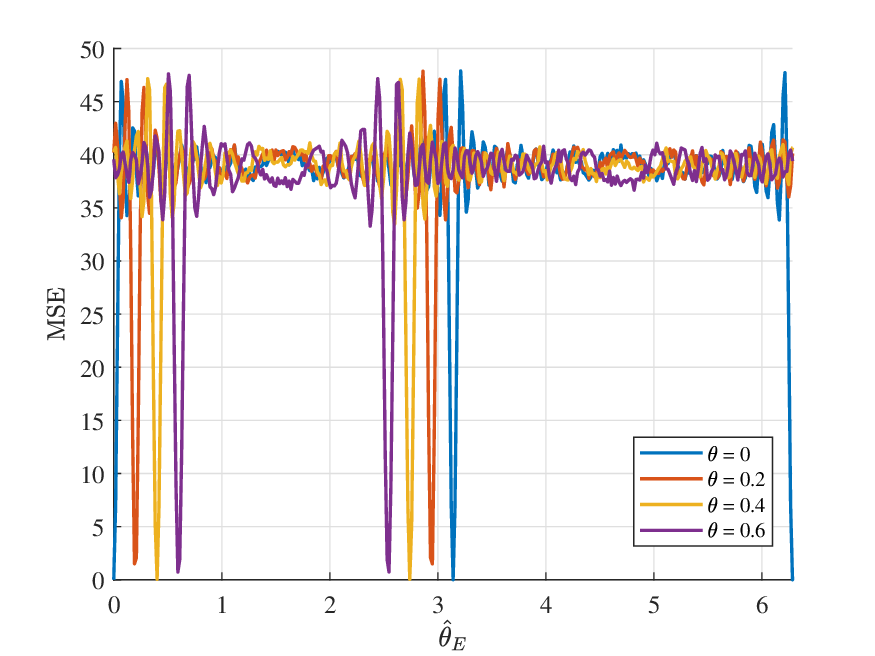}
\caption{MSE vs $\hat{\theta}_E$. Phases of Eve's precoding factors $\phi_1 = \phi_0 = 0$, amplitudes of Eve's precoding factors $\beta_0 + \beta_1 = 1$. $\textsc{SNR}_A = \textsc{SNR}_E$ = 30 dB. Number of Bob's antennas $M = 20$, number of Eve's antennas $L = 12$.}
\label{fig:MSE_vs_thetaE}
\par\end{centering}
\end{figure}

Fig. \ref{fig:MSE_vs_thetaE} shows what happens to MSE when Eve's AoA varies. In the simulations, we considered $M =20$ antennas at Bob and $L=12$ antennas at Eve. The MSE achieves the minimum in positions corresponding to values equal to $\theta - \pi$. These values have been highlighted for the red curve, which corresponds to $\theta = 0.2$ \SI{}{rad}.
Note that when $\sin\hat{\theta}_0=\sin{\theta}$ at $\hat{\theta}_0=\theta$ or $\hat{\theta}_0=\pi-\theta$ in the range $\lbrack 0,\pi\rbrack$, and the results repeats every $2\pi$.
These results therefore confirm that an attack is feasible only if the adversary is in the same direction as the legitimate node. 

\begin{figure}[!ht]
\begin{centering}
   \includegraphics[width=0.45\textwidth]{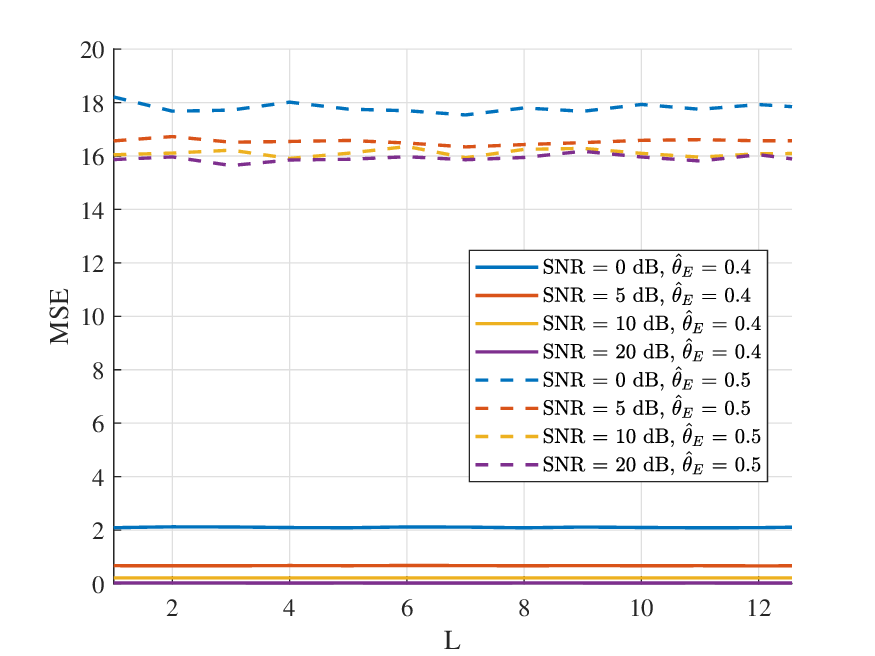}
\caption{MSE vs $L$. Phases of Eve's precoding factors $\phi_0 = \cdots = \phi_N = 0$, sum of amplitudes of Eve's precoding factors $\sum_{i}^{L-1}\beta_i = 1$. Alice's AoA $\theta = 0.4$ \SI{}{rad}.}
\label{fig:mse_vs_l}
\par\end{centering}
\end{figure}

Finally, we investigate the impact of the dimension of Eve's transmitting antenna array $L$ on the authentication. We suppose that Bob has 10 receiving antennas and that the SNR is the same on both legitimate channel and adversary channel. All the opponent's antennas are along the same direction, which we previously demonstrated to be the best possible attack scenario.
In Fig. \ref{fig:mse_vs_l} two conditions are investigated: the attacker can or cannot be in the same direction as Alice. It is possible to note that in the latter case, a degradation of performance (higher MSE) is clearly visible, as proved also in the previous simulations. Most importantly, the parameter $L$ has no clear impact on the results. Whether the attacker has fewer, the same number, or more antennas than the receiver, the MSE remains constant. Some little variation can be noticed when the SNR is low, but the overall MSE trend remains static. This result holds true even for higher numbers of antennas, which are not shown in the figure for readability reasons.

\subsection{Discussion}

Through analytical derivation and numerical results, we have demonstrated that AoA is a robust feature for authentication, even when an attacker has access to a large array of antennas. However, the system has a critical vulnerability: if all of the adversary's antennas align with the direction of the legitimate user's antennas, authentication can be compromised. In the following discussion, we explore potential methods to address this issue.

\noindent \textit{2D MUSIC}: A natural approach to enhance the security of the system using the same feature is to increase the number of AoA to predict by either using an antenna array at Bob and 2D MUSIC estimation or a distributed antenna system at Alice. 
Unfortunately, both schemes inherit the same key weakness as the present scheme - if an attacker knows the angles and places the antenna parallel with the propagation rays, it still has the chance to falsify the system, due to the fact that an angle alone is not enough to uniquely identify a location either in 2D or 3D systems. Another possibility is to place several base stations, which surely uniquely identify the location of the legitimate user. However, the base stations should be coordinated, otherwise, an attacker can still pose impersonation attacks utilizing a distributed antenna system, once it knows the corresponding angles. 

\begin{figure}
    \centering
    \includegraphics[width=0.7\linewidth]{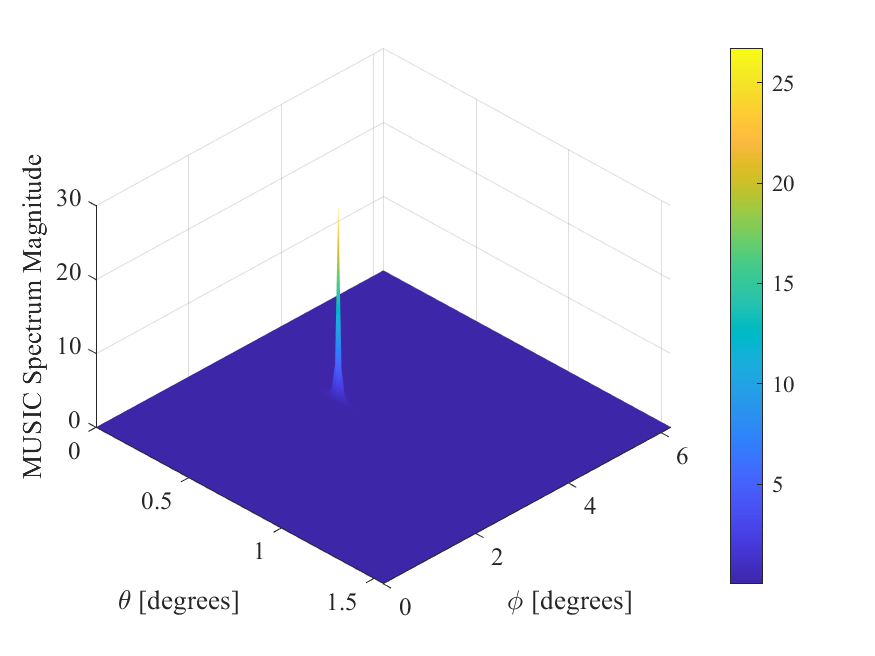}
    \caption{Output of 2D MUSIC algorithm for an input SNR of 0 dB.}
    \label{fig:2DMUSIC}
\end{figure}

As an example we demonstrate in Fig. \ref{fig:2DMUSIC}  the output of the 2D MUSIC algorithm which allows to uniquely identify a point \textit{on a plane}, not in a space. 
However, 2D MUSIC algorithms that also perform ranging have been proposed \cite{3DMUSIC}, allowing to uniquely identify a target in space.

Building on the contributions of the present work, AoA is a robust method to authenticate the location of a device. 

\noindent \textit{Multi-factor authentication}: AoA-based PLA is expected to perform well in stationary indoor scenarios. However, to make the authentication protocol strong and secure, we can leverage several methods to identify legitimate devices. For instance, we can combine AoA-based PLA with any of cryptography-based authentication, physical unclonable functions (PUFs), or RF-fingerprinting, proximity estimation in which the choice of attributes will depend on the application. Doing so, we may also increase the reliability of the authentication as a result of the exploitation of many authentication factors. We also note that for a certain channel, we can extract it into two components: static and dynamic ones. We can then utilize the former together with AoA for authentication and the latter for secret key generation \cite{BITS, Murali23}.

\section{Applications}

In this section we list and describe some possible applications of AoA-based physical layer authentication.

\subsection{Continuous Authentication}
 
In the traditional authentication approach, a user is authenticated at the beginning of a session. However, an attack could occur after authentication has been completed, and compromise an existing session.
Thus, the idea of continuous authentication \cite{Shepherd1995} is proposed to passively re-authenticate users without notifying or requesting special actions from them during the session.
Continuous authentication, in fact, provides an implicit means of user identity confirmation in an ongoing process, and usually leverages behavioral biometrics as identity information \cite{Patel2016, Quintal2019}.
Continuous authentication is also related to implicit authentication\cite{Jakobsson2009}, which is a form of authentication performed on the basis of the normal functions carried out by the device on behalf of the user, without the need to perform specific procedures for authentication.
AoA-based authentication can be used to provide two-factor-alike continuous authentication, to complement existing authentication schemes in the upper layers. The overall authentication mechanisms can work as follows.

\subsubsection{Enrollment phase}
 
\begin{enumerate}[i)]
\item Alice sends a data transmission request to Bob including some upper-layer authentication features, for example, a cryptographic-based one.
\item Bob receives the request and checks the legitimacy of the message by using the associated mechanism, for instance a cryptographic one. 
\item If the legitimacy of the user is confirmed, Bob performs AoA estimation and adds the estimated AoA feature to an access control list.  Otherwise, the request is rejected.
 \end{enumerate}

\subsubsection{Authentication phase}

\begin{enumerate}[i)]
\item Alice sends data to Bob using the same setting/channel used in the training phase.
\item Bob receives the signal and performs AoA estimation, comparing the estimated feature with that in the access control list. 
\item If the legitimacy of the user is confirmed, Bob accepts the signal and decodes it for further processing.  Otherwise, the signal is discarded.
\end{enumerate}

Clearly, the enrollment phase involves acquiring the specific AoA of the user to be authenticated, which serves to ensure that he or she is not subject to impersonation attacks as long as the relative position of the two users does not change.


   \subsection{Automatic Device Enrollment}


Another potential application of AoA-based PLA lies in zero-touch automation, particularly in automated device enrollment (ADE). With AoA-based PLA, we can envision smart environments such as factories or hospitals featuring designated \textit{authenticated locations}, where any device present is automatically recognized as legitimate. To achieve this, the access points must be configured during the initial setup to learn the AoA values corresponding to these authenticated locations. Once trained, the access points can verify the legitimacy of devices positioned in those locations, eliminating the need for higher-layer authentication protocols.




\subsection{Physical Layer Authentication and Key Agreement}

One further application of AoA-based PLA could be in conjunction with secret key generation (SKG) in realizing a physical layer procedure for authentication and key agreement (AKA).
This solution may be particularly promising for providing fast and lightweight AKA for delay-constrained scenarios, e.g., for a smart factory. Such protocols are usually referred to as PHY-AKA.

As an example, we can assume a smart factory setting in which both low-end, resource-constrained devices and high-end devices (such as autonomous robots) are present. The devices are connected to any of multiple access points equipped with massive multiple input multiple output (mMIMO) arrays, and can also engage into direct machine-to-machine (M2M) type communication, under severe delay constraints (e.g., $\leq 1$ ms).
As for the suitability of AoA-based PLA jointly used with SKG under very aggressive delay constraints, the following considerations are in order:
\begin{itemize}
    \item For PLA based on AoA estimation, recent variants of MUSIC algorithm can achieve remarkable performance. As an example, for a $100$ antenna element automotive radar mMIMO scenario, wall-clock runtimes of less than $0.4$ ms have been reported in \cite{fastMUSIC22}. 
\item For fast SKG, it has been shown that the use of neural networks can decisively speed up privacy amplification, aided by an enhanced feature space including physical attributes of the environment (LoS or NLoS, slow fading vs fast fading, etc.). As an example, for the experimental SKG campaign in \cite{Mayya23} with a symbol period of $T_s = 17.1875 \ \mu$s, a $K=16$ filter-bank along with a $Q=3$-bit quantizer, suffices for real-time estimation of the hashing rate in less than $0.2$ ms with an LSTM network.
\end{itemize}

In such a setting, PLA and SKG can be jointly employed for PHY-AKA between any device and a given access point, as well as for device pairing in M2M communications. 

\section{Conclusions}
\label{sec:con}

In this paper we have proposed the use of the angle of arrival as a robust feature to implement physical layer authentication in a multiple-antenna scenario, where users are equipped with digital antenna arrays. We have proven analytically that an impersonation attack can be successfully carried out only under very stringent conditions, regardless of the dimension of the antenna array available to the adversary. These findings have been corroborated by numerical results, obtained by simulating different attack scenarios.

\appendices
\section{Proof of two antennas case}
\label{app:2ants}
From the definitions, we have 
\begin{equation}
\mathbf{a}^{H}\mathbf{a}=M
\end{equation}

\begin{equation}
\mathbf{a}^{H}\mathbf{\hat{\mathbf{A}}}\mathbf{\mathbf{\hat{q}}}
= b_{0}\hat{q}_{0}+b_{1}\hat{q}_{1}
\end{equation}

\begin{equation}
\mathbf{\mathbf{\hat{q}}}^{H}\hat{\mathbf{A}}^{H}\mathbf{a} 
  =c_{0}\hat{q}_{0}^{*}+c_{1}\hat{q}_{1}^{*}
\end{equation}

\begin{equation}
\mathbf{\mathbf{\hat{q}}}^{H}\hat{\mathbf{A}}^{H}\mathbf{\hat{\mathbf{A}}}\mathbf{\mathbf{\hat{q}}}  =(|\hat{q}_{0}|^{2}+|\hat{q}_{1}|^{2})M+\hat{q}_{0}\hat{q}_{1}^{*}d_{0}+\hat{q}_{0}^{*}\hat{q}_{1}d_{1}.
\end{equation}

We can thus obtain

\begin{align}
\zeta&=M-b_{0}\hat{q}_{0}-b_{1}\hat{q}_{1}-c_{0}\hat{q}_{0}^{*}-c_{1}\hat{q}_{1}^{*}+(|\hat{q}_{0}|^{2}+|\hat{q}_{1}|^{2})M \nonumber\\ &+\hat{q}_{0}\hat{q}_{1}^{*}d_{0}+\hat{q}_{0}^{*}\hat{q}_{1}d_{1} + \frac{1}{\delta_{n}}+\frac{1}{\delta_{\hat{n}}}.
\label{eq:mse}
\end{align}
Assume $\hat{q}_{0}=\beta_{0}e^{j\phi_{0}}$, and $\hat{q}_{1}=\beta_{1}e^{j\phi_{1}}$
Similar to the first scenario, we also consider two cases, in which the AoA of the adversary is equal or not equal to that of the 
legitimate transmitter. 

\textbf{Case 1} In the special case $\hat{\theta}_{0}=\hat{\theta}_{1}=\theta$, we obtain 
\begin{align}
\zeta&=M(1-2\beta_{0}\cos\phi_{0}-2\beta_{1}\cos\phi_{1}+\beta_{0}^{2}+\beta_{1}^{2}+2\beta_{0}\beta_{1})\nonumber\\&+\frac{1}{\delta_{n}}+\frac{1}{\delta_{\hat{n}}}.
\end{align}

Equivalently, we can rewrite the equation above as 

\begin{equation}
\zeta=M\left((\beta_{0}+\beta_{1}-\cos\phi_{0})^{2}+\sin^{2}\phi_{0}\right)+\frac{1}{\delta_{n}}+\frac{1}{\delta_{\hat{n}}}
\end{equation}
which only achieve the minimum at $\phi_{0}=0$ and $\beta_{0}+\beta_{1}=1$. In other words, $\hat{q}_{0}+\hat{q}_{1}=1$ and $\hat{q}_{0}, \hat{q}_{1}$ are real.
This is also the best scenario for the adversary, in which MSE can reach the global minimum, like the single-antenna case. 

\textbf{Case 2} The AoA of the adversary is not equal to that of the legitimate transmitter. 

Note that from the definition, we can calculate all relevant parameters as follows:

\begin{align}
b_{0}&=1+e^{j\kappa(\sin\theta-\sin\hat{\theta}_{0})}+\ldots+e^{j(M-1)\kappa(\sin\theta-\sin\hat{\theta}_{0})}\\ &=\frac{\sin(\frac{M}{2}(\sin\theta-\sin\hat{\theta}_{0}))}{\sin(\frac{1}{2}(\sin\theta-\sin\hat{\theta}_{0}))}e^{j\frac{M-1}{2}(\sin\theta-\sin\hat{\theta}_{0})}.
\end{align}

Similarly, we achieve

\begin{equation}
b_{1}=\frac{\sin(\frac{M}{2}(\sin\theta-\sin\hat{\theta}_{1}))}{\sin(\frac{1}{2}(\sin\theta-\sin\hat{\theta}_{1}))}e^{j\frac{M-1}{2}(\sin\theta-\sin\hat{\theta}_{1})}
\end{equation}

\begin{equation}
c_{0}=\frac{\sin(\frac{M}{2}(\sin\theta-\sin\hat{\theta}_{0}))}{\sin(\frac{1}{2}(\sin\theta-\sin\hat{\theta}_{0}))}e^{-j\frac{M-1}{2}(\sin\theta-\sin\hat{\theta}_{0})}
\end{equation}

\begin{equation}
c_{1}=\frac{\sin(\frac{M}{2}(\sin\theta-\sin\hat{\theta}_{1}))}{\sin(\frac{1}{2}(\sin\theta-\sin\hat{\theta}_{1}))}e^{-j\frac{M-1}{2}(\sin\theta-\sin\hat{\theta}_{1})}
\end{equation}

\begin{equation}
d_{0}=\frac{\sin(\frac{M}{2}(\sin\hat{\theta}_{0}-\sin\hat{\theta}_{1}))}{\sin(\frac{1}{2}(\sin\hat{\theta}_{0}-\sin\hat{\theta}_{1}))}e^{-j\frac{M-1}{2}(\sin\hat{\theta}_{0}-\sin\hat{\theta}_{1})}
\end{equation}

\begin{equation}
d_{1}=\frac{\sin(\frac{M}{2}(\sin\hat{\theta}_{0}-\sin\hat{\theta}_{1}))}{\sin(\frac{1}{2}(\sin\hat{\theta}_{0}-\sin\hat{\theta}_{1}))}e^{j\frac{M-1}{2}(\sin\hat{\theta}_{0}-\sin\hat{\theta}_{1})}.
\end{equation}

To find the optimums, we need to compute the derivatives, i.e.,  

\begin{align}
\frac{\partial\zeta}{\partial\beta_{0}}&=-b_{0}e^{j\phi_{0}}-c_{0}e^{-j\phi_{0}}+2\beta_{0}M\nonumber\\&+\beta_{1}d_{0}e^{j(\phi_{0}-\phi_{1})}+\beta_{1}d_{1}e^{-j(\phi_{0}-\phi_{1})}
\end{align}

\begin{align}
\frac{\partial\zeta}{\partial\phi_{0}}&=-j(b_{0}\beta_{0}e^{j\phi_{0}}+c_{0}\beta_{0}e^{-j\phi_{0}})\nonumber\\&+j(\beta_{0}\beta_{1}d_{0}e^{j(\phi_{0}-\phi_{1})}-\beta_{0}\beta_{1}d_{1}e^{-j(\phi_{0}-\phi_{1})})
\end{align}

\begin{align}
\frac{\partial\zeta}{\partial\beta_{1}}&=-b_{1}e^{j\phi_{1}}-c_{1}e^{-j\phi_{1}}+2\beta_{1}M\nonumber\\&+\beta_{0}d_{0}e^{j(\phi_{0}-\phi_{1})}+\beta_{0}d_{1}e^{-j(\phi_{0}-\phi_{1})}
\end{align}

\begin{align}
\frac{\partial\zeta}{\partial\phi_{1}}&=-j(b_{1}\beta_{1}e^{j\phi_{1}}-c_{1}\beta_{1}e^{-j\phi_{1}})\nonumber\\&-j(\beta_{0}\beta_{1}d_{0}e^{j(\phi_{0}-\phi_{1})}-\beta_{0}\beta_{1}d_{1}e^{-j(\phi_{0}-\phi_{1})}).
\end{align}

To find the stationary points, we can set 

\begin{align}
0&=-b_{0}e^{j\phi_{0}}-c_{0}e^{-j\phi_{0}}+2\beta_{0}M\nonumber\\&+\beta_{1}d_{0}e^{j(\phi_{0}-\phi_{1})}+\beta_{1}d_{1}e^{-j(\phi_{0}-\phi_{1})}\label{eq:first:cond}
\end{align}

\begin{align}
0&=b_{0}\beta_{0}e^{j\phi_{0}}-c_{0}\beta_{0}e^{-j\phi_{0}}\nonumber\\&-\beta_{0}\beta_{1}d_{0}e^{j(\phi_{0}-\phi_{1})}+\beta_{0}\beta_{1}d_{1}e^{-j(\phi_{0}-\phi_{1})}\label{eq:second:cond}
\end{align}

\begin{align}
0&=-b_{1}e^{j\phi_{1}}-c_{1}e^{-j\phi_{1}}+2\beta_{1}M\nonumber\\&+\beta_{0}d_{0}e^{j(\phi_{0}-\phi_{1})}+\beta_{0}d_{1}e^{-j(\phi_{0}-\phi_{1})}\label{eq:third:cond}
\end{align}

\begin{align}
0&=b_{1}\beta_{1}e^{j\phi_{1}}-c_{1}\beta_{1}e^{-j\phi_{1}}\nonumber\\&+\beta_{0}\beta_{1}d_{0}e^{j(\phi_{0}-\phi_{1})}-\beta_{0}\beta_{1}d_{1}e^{-j(\phi_{0}-\phi_{1})}\label{eq:fourth:cond}.
\end{align}

Sum up \eqref{eq:first:cond} and \eqref{eq:second:cond}, we obtain

\begin{equation}
-c_{o}+\beta_{1}d_{1}e^{j\phi_{1}}+\beta_{0}Me^{j\phi_{0}}=0.\label{eq:first:trans}
\end{equation}

Subtracting \eqref{eq:first:cond} and \eqref{eq:second:cond} results in

\begin{equation}
-b_{0}e^{j(\phi_{0}+\phi_{1})}+\beta_{1}d_{0}e^{j\phi_{0}}+\beta_{0}Me^{j\phi_{1}}=0.\label{eq:second:trans}
\end{equation}

Continue in the same fashion for \eqref{eq:third:cond} and \eqref{eq:fourth:cond},
we obtain

\begin{align}
-c_{1}+\beta_{0}d_{0}e^{j\phi_{0}}+\beta_{1}Me^{j\phi_{1}} & =0\label{eq:third:trans}\\
-b_{1}e^{j(\phi_{0}+\phi_{1})}+\beta_{0}d_{1}e^{j\phi_{1}}+\beta_{1}Me^{j\phi_{0}} & =0.\label{eq:fourth:trans}
\end{align}

Sum up \eqref{eq:first:trans} and \eqref{eq:fourth:trans}, we arrive
at

\begin{equation}
-c_{0}-b_{1}e^{j(\phi_{0}+\phi_{1})}+(\beta_{0}+\beta_{1})(d_{1}e^{j\phi_{1}}+Me^{j\phi_{0}})=0.
\end{equation}

In other words 
\begin{equation}
\beta_{0}+\beta_{1} =\frac{c_{0}+b_{1}e^{j(\phi_{0}+\phi_{1})}}{d_{1}e^{j\phi_{1}}+Me^{j\phi_{0}}}
 =\frac{c_{0}e^{-j\phi_{0}}+b_{1}e^{j\phi_{1}}}{M+d_{1}e^{j(\phi_{1}-\phi_{0})}} = \frac{z_{1}}{z_{2}}
 \end{equation}
 where
 \begin{align}
 z_{1} &=\frac{\sin(\frac{M}{2}(\sin\theta-\sin\hat{\theta}_{0}))}{\sin(\frac{1}{2}(\sin\theta-\sin\hat{\theta}_{0}))}e^{j(-\frac{M-1}{2}(\sin\theta-\sin\hat{\theta}_{0})-\phi_{0})}\\ &+\frac{\sin(\frac{M}{2}(\sin\theta-\sin\hat{\theta}_{1}))}{\sin(\frac{1}{2}(\sin\theta-\sin\hat{\theta}_{1}))}e^{j(\frac{M-1}{2}(\sin\theta-\sin\hat{\theta}_{1})+\phi_{1})}
 \end{align}
 \begin{align}
 z_{2}&=M\nonumber\\&+\frac{\sin(\frac{M}{2}(\sin\hat{\theta}_{0}-\sin\hat{\theta}_{1}))}{\sin(\frac{1}{2}(\sin\hat{\theta}_{0}-\sin\hat{\theta}_{1}))}e^{j(\frac{M-1}{2}(\sin\hat{\theta}_{0}-\sin\hat{\theta}_{1})+\phi_{1}-\phi_{0})}.
\end{align}

Since $\beta_{0}+\beta_{1}$ is real, thus we have

\begin{align}
-\frac{M-1}{2}(\sin\theta-\sin\hat{\theta}_{0})-\phi_{0} & =0\\
\frac{M-1}{2}(\sin\theta-\sin\hat{\theta}_{1})+\phi_{1} & =0\\
\frac{M-1}{2}(\sin\hat{\theta}_{0}-\sin\hat{\theta}_{1})+\phi_{1}-\phi_{0} & =0.
\end{align}

It is easy to see that the solutions to the aforementioned equations are given by

\begin{equation}
\hat{\theta}_{0}=\hat{\theta}_{1}
\end{equation}

and 

\begin{equation}
\phi_{1}=\phi_{0}.
\end{equation}

As a result, we obtain $\beta_{1}+\beta_{0}=\frac{1}{M}\frac{\sin(\frac{M}{2}(\sin\theta-\sin\hat{\theta}_{0}))}{\sin(\frac{1}{2}(\sin\theta-\sin\hat{\theta}_{0}))}$.

In summary, we achieve the following to minimize the MSE in case the AoAs of the legitimate and the adversary are different 

\begin{align}
\hat{\theta}_{0} & =\hat{\theta}_{1}\\
\phi_{0} & =\phi_{1}
\end{align}

\begin{equation}
\beta_{1}+\beta_{0}=\frac{1}{M}\frac{\sin(\frac{M}{2}(\sin\theta-\sin\hat{\theta}_{0}))}{\sin(\frac{1}{2}(\sin\theta-\sin\hat{\theta}_{0}))}.
\end{equation}
As a consequence of the preceding  proofs, $\zeta$ only achieve the global minimum at $\hat{q}_{0}+\hat{q}_{1}=1$ and $\hat{q}_{0}, \hat{q}_{1}$ are real, which concludes the proof.

\balance
\bibliographystyle{IEEEtran}
\bibliography{bib}

\begin{thebibliography}{10}
\providecommand{\url}[1]{#1}
\csname url@samestyle\endcsname
\providecommand{\newblock}{\relax}
\providecommand{\bibinfo}[2]{#2}
\providecommand{\BIBentrySTDinterwordspacing}{\spaceskip=0pt\relax}
\providecommand{\BIBentryALTinterwordstretchfactor}{4}
\providecommand{\BIBentryALTinterwordspacing}{\spaceskip=\fontdimen2\font plus
\BIBentryALTinterwordstretchfactor\fontdimen3\font minus
  \fontdimen4\font\relax}
\providecommand{\BIBforeignlanguage}[2]{{%
\expandafter\ifx\csname l@#1\endcsname\relax
\typeout{** WARNING: IEEEtran.bst: No hyphenation pattern has been}%
\typeout{** loaded for the language `#1'. Using the pattern for}%
\typeout{** the default language instead.}%
\else
\language=\csname l@#1\endcsname
\fi
#2}}
\providecommand{\BIBdecl}{\relax}
\BIBdecl

\bibitem{Pham2023}
T.~M. Pham, L.~Senigagliesi, M.~Baldi, G.~P. Fettweis, and A.~Chorti, ``Machine
  learning-based robust physical layer authentication using angle of arrival
  estimation,'' in \emph{Proc. IEEE GLOBECOM}, 2023, pp. 13--18.

\bibitem{Chorti:Contextaware:PLS:2022}
A.~Chorti, A.~N. Barreto, S.~Köpsell, M.~Zoli, M.~Chafii, P.~Sehier,
  G.~Fettweis, and H.~V. Poor, ``Context-aware security for 6{G} wireless: The
  role of physical layer security,'' \emph{IEEE Commun. Stand. Mag.}, vol.~6,
  no.~1, pp. 102--108, 2022.

\bibitem{Chorti23}
M.~Mitev, A.~Chorti, H.~V. Poor, and G.~P. Fettweis, ``What physical layer
  security can do for {6G} security,'' \emph{IEEE Open J. Veh. Technol.},
  vol.~4, pp. 375--388, 2023.

\bibitem{Mitev22}
M.~Mitev, M.~Shakiba-Herfeh, A.~Chorti, M.~Reed, and S.~Baghaee, ``A physical
  layer, zero-round-trip-time, multifactor authentication protocol,''
  \emph{IEEE Access}, vol.~10, pp. 74\,555--74\,571, 2022.

\bibitem{Pappu:PUF:2022}
R.~Pappu, B.~Recht, J.~Taylor, and N.~Gershenfeld, ``Physical one-way
  functions,'' \emph{Science}, vol. 297, no. 5589, pp. 2026--2030, 2002.

\bibitem{Gassend:SilionPUF:2002}
B.~Gassend, D.~Clarke, M.~van Dijk, and S.~Devadas, ``Silicon physical random
  functions.''\hskip 1em plus 0.5em minus 0.4em\relax New York, NY, USA:
  Association for Computing Machinery, 2002.

\bibitem{Hao:IQauthentication:2014}
P.~Hao, X.~Wang, and A.~Behnad, ``Performance enhancement of i/q imbalance
  based wireless device authentication through collaboration of multiple
  receivers,'' in \emph{Proc. IEEE ICC}, 2014, pp. 939--944.

\bibitem{Hao:RelayIQ:2014}
------, ``Relay authentication by exploiting i/q imbalance in
  amplify-and-forward system,'' in \emph{Proc. IEEE GLOBECOM}, 2014, pp.
  613--618.

\bibitem{Chorti22}
M.~Srinivasan, S.~Skaperas, M.~S. Herfeh, and A.~Chorti, ``Joint
  localization-based node authentication and secret key generation,'' in
  \emph{Proc. IEEE ICC}, 2022, pp. 32--37.

\bibitem{Passah24}
A.~K.~A. Passah, A.~Chorti, and R.~C. de~Lamare, ``Enhanced multiuser
  {CSI}-based physical layer authentication based on information
  reconciliation,'' \emph{IEEE Wirel. Commun.}, pp. 1--1, 2024.

\bibitem{Faria:Signalprints:2006}
D.~B. Faria and D.~R. Cheriton, ``Detecting identity-based attacks in wireless
  networks using signalprints,'' ser. WiSe '06.\hskip 1em plus 0.5em minus
  0.4em\relax New York, NY, USA: Association for Computing Machinery, 2006, p.
  43–52.

\bibitem{Xiao:EnhanceAuthentication:2008}
L.~Xiao, L.~Greenstein, N.~Mandayam, and W.~Trappe, ``A physical-layer
  technique to enhance authentication for mobile terminals,'' in \emph{Proc.
  IEEE ICC}, 2008, pp. 1520--1524.

\bibitem{Liu:CIR:PLA:2011}
F.~J. Liu, X.~Wang, and H.~Tang, ``Robust physical layer authentication using
  inherent properties of channel impulse response,'' in \emph{Proc. IEEE
  MILCOM}, 2011, pp. 538--542.

\bibitem{Liu:2Dquantization:PLA:2013}
F.~J. Liu, X.~Wang, and S.~L. Primak, ``A two dimensional quantization
  algorithm for cir-based physical layer authentication,'' in \emph{Proc. IEEE
  ICC}, 2013, pp. 4724--4728.

\bibitem{Xiao:CFR:PLA:2008}
L.~Xiao, L.~J. Greenstein, N.~B. Mandayam, and W.~Trappe, ``Using the physical
  layer for wireless authentication in time-variant channels,'' \emph{IEEE
  Wirel. Commun.}, vol.~7, no.~7, pp. 2571--2579, 2008.

\bibitem{Senigagliesi21}
L.~Senigagliesi, M.~Baldi, and E.~Gambi, ``Comparison of statistical and
  machine learning techniques for physical layer authentication,'' \emph{IEEE
  Trans. Inf. Forensics Secur.}, vol.~16, pp. 1506--1521, 2021.

\bibitem{Xiong:Securearray:2013}
J.~Xiong and K.~Jamieson, ``Secure{A}rray: Improving wifi security with
  fine-grained physical-layer information,'' in \emph{Proc. ACM MobiCom}, ser.
  MobiCom '13.\hskip 1em plus 0.5em minus 0.4em\relax New York, NY, USA:
  Association for Computing Machinery, 2013, p. 441–452.

\bibitem{Abdelaziz2019}
A.~Abdelaziz, R.~Burton, F.~Barickman, J.~Martin, J.~Weston, and C.~E. Koksal,
  ``Enhanced authentication based on angle of signal arrivals,'' \emph{IEEE
  Trans. Veh. Technol.}, vol.~68, no.~5, pp. 4602--4614, 2019.

\bibitem{Xu2022}
W.~Xu, L.~Tao, and Q.~Xu, ``Physical layer authentication based on {DOA} and
  rotational state,'' in \emph{Proc. IEEE WCSP}, 2022, pp. 1028--1033.

\bibitem{Topal2022}
O.~A. Topal and G.~Karabulut~Kurt, ``Physical layer authentication for {LEO}
  satellite constellations,'' in \emph{Proc. IEEE WCNC}, 2022, pp. 1952--1957.

\bibitem{Casari2022}
P.~Casari, F.~Ardizzon, and S.~Tomasin, ``Physical layer authentication in
  underwater acoustic networks with mobile devices,'' in \emph{Proc. ACM
  WUWNet}, 2022, pp. 1--8.

\bibitem{Li:Proofing:Channel:AoA:2021}
W.~Li, N.~Wang, L.~Jiao, and K.~Zeng, ``Physical layer spoofing attack
  detection in mmwave massive {MIMO 5G} networks,'' \emph{IEEE Access}, vol.~9,
  pp. 60\,419--60\,432, 2021.

\bibitem{Abdelaziz:Sec:AoA:2016}
A.~Abdelaziz, C.~E. Koksal, and H.~El~Gamal, ``On the security of angle of
  arrival estimation,'' in \emph{Proc. IEEE CNS}, 2016, pp. 109--117.

\bibitem{Murali24}
M.~Srinivasan, L.~Senigagliesi, H.~Chen, A.~Chorti, M.~Baldi, and H.~Wymeersch,
  ``Aoa-based physical layer authentication in analog arrays under
  impersonation attacks,'' in \emph{Proc. IEEE SPAWC}, 2024, pp. 496--500.

\bibitem{Godara:DS:1997}
L.~Godara, ``Application of antenna arrays to mobile communications. {II}.
  {B}eam-forming and direction-of-arrival considerations,'' \emph{Proc. IEEE},
  vol.~85, no.~8, pp. 1195--1245, 1997.

\bibitem{Capon:MDVR:1969}
J.~Capon, ``High-resolution frequency-wavenumber spectrum analysis,''
  \emph{Proc. IEEE}, vol.~57, no.~8, pp. 1408--1418, 1969.

\bibitem{Schmidt:MUSIC:1986}
R.~Schmidt, ``Multiple emitter location and signal parameter estimation,''
  \emph{IEEE Trans. Antennas Propag.}, vol.~34, no.~3, pp. 276--280, 1986.

\bibitem{Samuel:Funfact:1986}
S.~{Greitzer}, ``Many cheerful facts,'' \emph{Arbelos 4}, no.~5, pp. 14--17,
  1986.

\bibitem{3DMUSIC}
E.~Zhao, F.~Zhang, D.~Zhang, and S.~Pan, ``Three-dimensional multiple signal
  classification (3d-music) for super-resolution fmcw radar detection,'' in
  \emph{Proc. IEEE IWS}, 2019, pp. 1--3.

\bibitem{BITS}
M.~Mitev, T.~M. Pham, A.~Chorti, A.~N. Barreto, and G.~Fettweis, ``Physical
  layer security—from theory to practice,'' \emph{IEEE BITS the Information
  Theory Magazine}, vol.~3, no.~2, pp. 67--79, 2023.

\bibitem{Murali23}
M.~Srinivasan, S.~Skaperas, M.~Mitev, M.~S. Herfeh, M.~K. Shehzad, P.~Sehier,
  and A.~Chorti, ``Smart channel state information pre-processing for
  authentication and symmetric key distillation,'' \emph{IEEE IEEE Trans. Mach.
  Learn. Commun. Netw.}, vol.~1, pp. 328--345, 2023.

\bibitem{Shepherd1995}
S.~Shepherd, ``Continuous authentication by analysis of keyboard typing
  characteristics,'' in \emph{European Convention on Security and Detection,
  1995.}, 1995, pp. 111--114.

\bibitem{Patel2016}
V.~M. Patel, R.~Chellappa, D.~Chandra, and B.~Barbello, ``Continuous user
  authentication on mobile devices: Recent progress and remaining challenges,''
  \emph{IEEE Signal Process. Mag.}, vol.~33, no.~4, pp. 49--61, 2016.

\bibitem{Quintal2019}
K.~Quintal, B.~Kantarci, M.~Erol-Kantarci, A.~Malton, and A.~Walenstein,
  ``Contextual, behavioral, and biometric signatures for continuous
  authentication,'' \emph{IEEE Internet Comput.}, vol.~23, no.~5, pp. 18--28,
  2019.

\bibitem{Jakobsson2009}
M.~Jakobsson, E.~Shi, P.~Golle, and R.~Chow, ``Implicit authentication for
  mobile devices,'' in \emph{HotSec 2009 - 4th USENIX Workshop on Hot Topics in
  Security}, 2009.

\bibitem{fastMUSIC22}
B.~Li, S.~Wang, J.~Zhang, X.~Cao, and C.~Zhao, ``Ultra-fast accurate aoa
  estimation via automotive massive-{MIMO} radar,'' \emph{IEEE Trans. Veh.
  Technol.}, vol.~71, no.~2, pp. 1172--1186, 2022.

\bibitem{Mayya23}
A.~Mayya, M.~Mitev, A.~Chorti, and G.~Fettweis, ``A {SKG} security challenge:
  Indoor {SKG} under an on-the-shoulder eavesdropping attack,'' in \emph{Proc.
  IEEE GLOBECOM}, 2023, pp. 3451--3456.

\end{thebibliography}

\end{document}